\documentclass[twocolumn,prb,aps,amsmath,amssymb,superscriptaddress,showpacs]{revtex4-2}
\usepackage{graphicx}
\usepackage{amsthm}
\usepackage{bm}
\usepackage[usenames,dvipsnames]{color}
\usepackage{tikz-cd}
\usepackage{hhline}

\renewcommand\Re{\operatorname{Re}}

\newtheorem{theorem}{Theorem}

\newcommand{\bfun}{\mathbf{1}}

\newcommand{\bbR}{\mathbb{R}}

\newcommand\e{\mathrm{e}}

\newcommand\ee{\e}

\newcommand\Tr{\mathrm{Tr}}

\newcommand\id{\mathrm{Id}}

\newcommand\calH{\mathcal{H}}

\newcommand\bfa{\mathbf{a}}

\newcommand\bfe{\mathbf{e}}

\newcommand\bfk{\mathbf{k}}

\newcommand\bfp{\mathbf{p}}
\newcommand\bfP{\mathbf{P}}

\newcommand\bfr{\mathbf{r}}

\newcommand\bfR{\mathbf{R}}
\newcommand\bfu{\mathbf{u}}
\newcommand\bfv{\mathbf{v}}

\newcommand{\bfpi}{{\boldsymbol{\pi}}}

\newcommand{\ninej}[9]{\left\{ \begin{array}{@{}c@{\;}c@{\;}c@{}}
     		#1 & #2 & #3 \\
     		#4 & #5 & #6 \\
       		#7 & #8 & #9
     	\end{array}\right\}}

\usepackage[version=4]{mhchem}
\newcommand{\tetraCl}{\ce{U^{VI}O2Cl4^{2-}}}
\newcommand{\UOsix}{\ce{Sr3U^{VI}O6}}
\newcommand{\UOtwo}{\ce{U^{IV}O2}}
\newcommand{\threeD}{$3d$ }
\newcommand{\fourF}{$4f$ }
\newcommand{\fiveF}{$5f$ }
\newcommand{\threeDfourF}{$3d4f$}
\newcommand{\threeDdashfiveF}{$3d$-$5f$}
\newcommand{\fourFdashfiveF}{$4f$-$5f$}
\newcommand{\fiveFdashfiveF}{$5f$-$5f$}
\newcommand{\DFourH}{$D_{4h}$}
\newcommand{\Oh}{$O_{h}$}
\newcommand{\ki}{$\widehat\bfk_i$}
\newcommand{\ks}{$\widehat\bfk_s$}
\usepackage{verbatim}
\usepackage{hyperref}
\usepackage{cleveref}

\usepackage{cancel}

\begin{document}
\title{Angular dependence and powder average of resonant inelastic X-ray scattering}

\author{Myrtille O. J. Y Hunault}
\author{Timothy G. Burrow} 
\affiliation{Synchrotron SOLEIL, L'Orme des Merisiers, 91190 Saint-Aubin, France}

\author{Fabien Besnard}
\affiliation{11 all\'ee Hector Berlioz, 95230 Soisy-sous-Montmorency, France}

\author{Am\'elie Juhin}
\email[Corresponding author: ] {amelie.juhin@sorbonne-universite.fr}
\author{Christian Brouder}
\affiliation{Institut de Min\'eralogie, de Physique des Mat\'eriaux et 
	de Cosmochimie, IMPMC, UMR CNRS 7590, Sorbonne Universit\'e, Mus\'eum National d'Histoire Naturelle, 
	  75005 Paris, France}

\date{\today}

\begin{abstract}

Resonant inelastic X-ray scattering (RIXS) is a synchrotron-based spectroscopy that has seen growing interest across a range of scientific disciplines beyond fundamental physics. The interpretation of experimental RIXS data requires theoretical calculations based on the Kramers–Heisenberg formula. However, due to the dependence of RIXS on both the incident and scattered photon properties, a tractable treatment of the angular dependence in this formula has been lacking.
In this work, within the electric dipole approximation, we determine the number of fundamental spectra contributing to the RIXS cross-section for all crystallographic point groups. We then derive a general expression for the RIXS cross-section of isotropic samples such as un-textured powders, homogeneous glasses or liquids, explicitly accounting for the polarization and propagation directions of both the incident and scattered photons. Simplified forms of the RIXS expressions are subsequently obtained for most common point groups. Finally, we demonstrate the applicability of our formalism through a case study of uranium $3d4f$ RIXS.
\end{abstract}

\maketitle

Resonant inelastic X-ray scattering (RIXS) has proven to be extremely useful to investigate the properties of materials \cite{Ament-11,de_groot_resonant_2024}.
The riches of RIXS come from the fact that it depends on the energy of the incident and scattered beams, as well as
on the polarization and direction of the incident and scattered beams with respect to the sample, which allows access to more spectral information \cite{ Ishii2011,Minola_2015,Inami_2017,Heba_2019}. 
To this end, instrumental capabilities have been implemented at synchrotron beamlines, providing the ability to control the polarization of the incident beam, vary the scattering angle, and perform polarization analysis of the scattered radiation \cite{brookes_beamline_2018,zhou_i21_2022,yamamoto_new_2014}.

While most of these studies are performed on single crystals, the number of studies on isotropic samples like powders has been growing thanks to the application of RIXS by new scientific communities, e.g. chemistry~\cite{jay_capturing_2022,de_groot_resonant_2024}, catalysis~\cite{cutsail_iii_challenges_2022}, high-pressure~\cite{rosa_new_2024}, actinides~\cite{prieur_frontiers_2025}. 
These communities exploit RIXS to go beyond conventional X-ray Absorption Spectroscopy (XAS) through the use of High Energy Resolution Fluorescence Detection (HERFD). This approach, which involves taking a constant-emission-energy cut in the RIXS plane, overcomes the limitations imposed by core-hole lifetime broadening and reveals otherwise hidden spectral features. A remarkable example is provided by the actinide M edges, where the application of HERFD-XAS has reopened a very active field of investigation on these complex compounds~\cite{kvashnina_invisible_2014,kvashnina_high-energy_2022}.
However, the interpretation of experimental RIXS data often requires theoretical calculations. The Kramers-Heisenberg formula~\cite{Kramers-25,Kramers-25-GB} has been implemented in various computational methods, from semi-empirical ligand field multiplet calculations~\cite{glatzel_high_2005} to more advanced \textit{ab-initio} calculations~\cite{kas_real-space_2011,maganas_restricted_2017}.

In a previous work~\cite{Juhin-14}, we investigated the
dependence of RIXS on the polarization and direction of the incident and scattered
beams for a sample without considering the effect of sample symmetry. We also derived explicit formulas
for the polarization and angular dependence of an isotropic powder sample. 
However, the remaining challenge is that computational methods do not use an isotropic sample as input but the non-isotropic structural model of the system of interest (e.g. the point group symmetry of the absorbing atom for semi-empirical multiplet calculations, a cluster around the absorber or the unit cell of a crystal for \textit{ab-initio} methods). In standard XAS computation, the spectra computed using the model system can easily be related to the experimental isotropic spectrum \cite{BrouderAXAS}. However for RIXS, the coupling between the incident photon and the scattered photon makes this connection more challenging.

The purpose of this article is two-fold. Firstly, we discuss the effect of a sample
(molecule or crystal) symmetry on the angular and polarization dependencies and secondly we
determine the minimal number of calculations (or single-crystal measurements) required
to compute these dependencies for an isotropic powder sample in the electric dipole approximation for the incident and scattered beams. These derivations are performed in the impurity approximation, meaning that we neglect the interferences between the scattering events of different atoms.
In the following, the coordinates $(x,y,z)$ are in the standard orthonormal reference frame for point group operations as defined by Altman and Herzig~\cite{Altmann}.

This paper is divided into six parts. In part I, we demonstrate the well-known fact that the RIXS cross-section 
for electric dipole transitions is based on 81 spectra, and we prove that this number can be reduced further down 
to $n_G\le 81$ fundamental spectra by taking symmetry into account. In part II, we use spherical tensors to understand $n_G$ and provide specific considerations about circular dichroism and site symmetry. In part III, we determine $n_G$ and the relations between RIXS matrix elements for several symmetry groups as well as for the case of a spherically symmetric sample. In part IV, we derive the RIXS cross-section formula for a powder sample without taking the symmetry into account and then in part V, we give the final RIXS cross-section formula for an isotropic sample knowing the symmetry of the absorber for most point groups. Finally, in part VI, we draw some predictions for the experimental observations and we apply these formulas to selected examples to highlight the impact on the computed result. 

\section{RIXS cross section and sample symmetry}

The starting point of this analysis is the
Kramers-Heisenberg formula \cite{Kramers-25,Kramers-25-GB}
for RIXS
\begin{eqnarray*}
	&& \frac{r_e^2}{m^2}
	\frac{\omega_{s}}{\omega}\sum_F
	\Big|
	\sum_N
	\frac{\langle F| \epsilon_s^*\cdot\bfP
		\ee^{-i\bfk_s\cdot\bfr} |N\rangle
		\langle N| \epsilon\cdot\bfP
		\ee^{i\bfk\cdot\bfr} |I\rangle}
	{E_I-E_N+\hbar\omega+i\gamma}
	\Big|^2 
	\\&&\times
	\delta(E_F+\hbar\omega_{s}-E_I-\hbar\omega),
\end{eqnarray*}
where
$m$ is the electron mass, $r_e$ is the classical electron radius:
$r_e=e^2/(4\pi\epsilon_0 mc^2)$,
$|I\rangle$, $|N\rangle$, $|F\rangle$ are the initial, intermediate and final 
(possibly many-body) states, respectively,
$\gamma$ is the total width of the intermediate state
$|N\rangle$, $\bfP$ and $\bfr$ are the momentum and position operators. The incident and scattered photons are characterized by the pulsation,
wavevector and polarization vectors $\omega,\bfk,\epsilon$ and
$\omega_s,\bfk_s,\epsilon_s$, respectively. Note that $\epsilon_s^*$
is the complex conjugate of $\epsilon_s$.

In the present paper we limit ourselves to 
the RIXS cross-section due to
electric dipole transitions~\cite{Juhin-14}:
\begin{eqnarray}
	\sigma(\epsilon_s,\epsilon)
	&=& \frac{r_e^2}{\hbar^2}\frac{\omega_{s}}{\omega}\sum_F
	\Big|
	\sum_N \frac{(E_I-E_N)(E_N-E_F)}{E_I-E_N+\hbar\omega+i\gamma} 
	\nonumber\\&&\times
	\langle F|\epsilon_s^*\cdot\bfr |N\rangle\langle N|\epsilon\cdot\bfr |I\rangle
	\Big|^2 \delta_{E}, 
	\label{scatcc2}
\end{eqnarray}
where $\delta_{E}=\delta(E_F+\hbar\omega_{s}-E_I-\hbar\omega)$.

\subsection{Group action}

It will be convenient to use the concept of an action of a group on a set.
Consider for example the group $G=O(3)$ of (linear) isometries, where
an element $g$ of $G$ is represented by an orthogonal matrix $g_{ij}$
(i.e. a rotation matrix possibly multiplied by an inversion). 
With $g$ we can operate on:
{\it {i)}} a vector $\bfr$, {\it {ii)}} a function $f(\bfr)$, {\it {iii)}} a
quantum state $|\phi\rangle$, {\it {iv)}} a sample $s$, etc. 

In general, the action of an operation $g$ on a set $X$ is a 
transformation of any element $x$ of $X$ into an element $y$ of $X$, 
denoted by $y=g\triangleright x$.
In our examples: {\it {i)}} $X$ is the space $\bbR^3$ of 3-dimensional vectors $\bfr$,
{\it {ii)}} $X$ is the space of functions $f(\bfr)$ on $\bbR^3$, {\it {iii)}} 
$X$ is the Hilbert space $\calH$ of quantum states
$|\phi\rangle$ of the system, {\it {iv)}} $X$ is
the set $S$ made of a sample $s$ with all its possible orientations
(note that we can rotate an object but we cannot 
mechanically invert it). 

An action of a group $G$ on a set $X$ is a \emph{group action} 
if the successive action of two arbitrary group elements is compatible with the
group product : $g'\triangleright(g\triangleright x)=(g'\cdot g)\triangleright x$,
where $g'\cdot g$ is the product in $G$,
and if all elements $x$ of $X$ are stable under the action of
the identity element 
$e$ of $G$ (i.e. $e\triangleright x=x$).
In case {\it {i)}}  the action of an isometry $g$
on a vector $\bfr$ is $g\triangleright \bfr=g\bfr$
(i.e. product of the matrix $g_{ij}$ by the vector $\bfr$).
In case {\it {ii)}}
the action of a rotation on a function $f(\bfr)$ is~\cite{Letouze-23}
\begin{eqnarray}
	(g\triangleright f)(\bfr)&=&f(g^{-1}\bfr),
	\label{Rtrianglef}
\end{eqnarray}
where $g^{-1}_{ij}$ is the inverse matrix of $g_{ij}$.
For case {\it {iii)}}, we consider the action of an isometry on a quantum state
which is an eigenstate of a Hamiltonian $H$ with energy $E$.
To take energy degeneracy into account, consider a
basis $|\psi_1\rangle,\dots,|\psi_n\rangle$ of the eigenstates of 
$H$ with energy $E$. 
If $G$ is a symmetry group of $H$ (i.e. $g\triangleright H=H$ for all
$g\in G$), we know that every element $g$ of $G$ transforms
a state of $H$ into a state with the same energy. 
More precisely, Wigner showed that there is a unitary matrix $U(g)$
such that:\cite{Wigner}
\begin{eqnarray}
	g\triangleright |\psi_i\rangle &=& \sum_{j=1}^n U(g)_{ji}|\psi_j\rangle,
	\label{gtrianglePsi}
\end{eqnarray}
where $U(g)_{ji}$ is required instead of $U(g)_{ij}$ for $g\triangleright |\psi_i\rangle$
 to be a (linear) group action:
\begin{eqnarray*}
	g'\triangleright(g\triangleright |\psi_i\rangle) &=& \sum_{j=1}^n U(g)_{ji}
	(g'\triangleright|\psi_j\rangle)
	\\&=&
	\sum_{j=1}^n\sum_{k=1}^n U(g)_{ji} U(g')_{kj}|\psi_k\rangle
	\\&=& \sum_{k=1}^n \big(U(g')U(g)\big)_{ki}|\psi_k\rangle.
\end{eqnarray*}
This action is a group action if and only if
$U(g')U(g)=U(g'\cdot g)$ (i.e. $U$ is a unitary representation of $G$).

Consider now the projection onto the space of states of energy $E$:
\begin{eqnarray*}
P(E) &=& \sum_{j=1}^n|\psi_j\rangle\langle\psi_j|.
\end{eqnarray*}
From the action of $g$ on $|\psi_j\rangle$ of Eq.~\eqref{gtrianglePsi}
and the unitarity of $U(g)$,
we deduce that $P(E)$ is invariant under the action of $g$:
\begin{eqnarray}
	g\triangleright P(E) &=& \sum_{j,k,l}U(g)_{kj}|\psi_k\rangle\langle\psi_l|U(g)_{lj}^*
	=P(E).
	\label{gtrianglePE}
\end{eqnarray}

These preliminaries enable us to calculate various actions
of groups of isometries on the RIXS cross-section.

\subsection{Action of an isometry on the cross-section}
We calculate the action of isometries
on the RIXS cross-section in two ways: by acting on the states seen as a function of
$\bfr$ as in Eq.~\eqref{Rtrianglef} and then by acting on the states
by a unitary representation as in  Eq.~\eqref{gtrianglePsi}.

We want to determine how the 
transition matrix element
$\langle \phi |H_I |\psi\rangle =
\int d\bfr \phi^*(\bfr) \epsilon\cdot\bfr \psi(\bfr)$ transforms
under the action of a general isometry $g$.
We show that 
\begin{eqnarray}
	g\triangleright\langle \phi |H_I |\psi\rangle &=&
	\int d\bfr  \phi^*(g^{-1}\bfr) \epsilon\cdot\bfr \psi(g^{-1}\bfr).
	\label{gtriangleTME}
\end{eqnarray}
This statement is not as trivial as it looks
because we must disentangle the action of $g$ on
$H_I$ from its action on the unperturbed Hamiltonian $H_0$.
First recall that the Hamiltonian of
an electron in an electromagnetic field is 
$H=(\bfp-e \bfa)^2/2m +V+e v$, where $\bfp=-i\hbar\nabla$,
$\bfa$ is the vector potential and $v$ the electric potential.
Time-dependent perturbation theory for 
an incident electromagnetic wave described by
the vector potential $\bfa\cos(\omega t)$ leads 
to the calculation of
matrix elements such as $\langle\phi|H_I|\psi\rangle$, where
$H_I=-e\bfp\cdot\bfa/m$ and
$|\phi\rangle$ and $|\psi\rangle$ are eigenstates of 
$H_0=\bfp^2/2m+V$~\cite{BrouderAXAS}.
Within this model, acting with isometry $g$ on the sample amounts to 
transforming the potential $V(\bfr)$ into the potential 
$V'(\bfr)=(g\triangleright V)(\bfr)=V(g^{-1}\bfr)$, as in Eq.~\eqref{Rtrianglef}.
The Hamiltonian of the transformed sample in the original reference frame
is $H_0'=\bfp^2/2m+V'$ and we go from the momentum transition operator $-e\bfp\cdot\bfa/m$
to the electric dipole one by using ${[}\bfr,H_0{]}={[}\bfr,H'_0{]}=(\hbar^2/m)\nabla$.
Therefore, $g$ does not act on the transition operator, which is 
still $\epsilon\cdot\bfr$. 

To determine the eigenstates of $H'_0$, we observe that a change of coordinates
$\bfr\to\bfr'=g^{-1}\bfr$ transforms $H_0$
into $g\triangleright H_0=(\bfp')^2/2m+V(\bfr')$, where
$\bfp'=g^{-1}\bfp$. The orthogonality of $g$ implies that
$(\bfp')^2=\bfp^2$ and $g\triangleright H_0=H'_0$.
Thus, the eigenstates of $H'_0$ are the eigenstates of 
$g\triangleright H_0$, which are
$\phi(\bfr')=\phi(g^{-1}\bfr)$. This completes the proof
of Eq.~\eqref{gtriangleTME}. Note that the result 
is still valid if $H_0$ contains a two-body Coulomb interaction.

Starting from Eq~\eqref{gtriangleTME}, we change variables
$\bfr=g\bfr'$ and notice that $d\bfr=d\bfr'$
($g$ being an orthogonal matrix, the Jacobian of the
transformation is $|J|=|\det g|=1$)
to replace a transformation of the sample by 
a transformation of the transition operator
and then a transformation of the polarization vector:
\begin{eqnarray*}
	g\triangleright\langle \phi |\epsilon\cdot\bfr |\psi\rangle &=&
	\langle \phi |\epsilon\cdot(g\bfr) |\psi\rangle
=
\langle \phi |(g^{-1}\epsilon)\cdot\bfr |\psi\rangle,
\end{eqnarray*}
because $g$, as a linear isometry, conserves scalar products. 
The physical interpretation of this result is that 
the action of an isometry on a sample gives the same
result as the action of the inverse isometry 
on the measurement device (represented by $\epsilon$).

Thus, the action of an isometry $g$ on the sample
transforms the cross-section into
$g\triangleright\sigma(\epsilon_s,\epsilon)=
	\sigma(g^{-1}\epsilon_s,g^{-1}\epsilon)$,
that it will be more convenient to write
\begin{eqnarray}
	g^{-1}\triangleright\sigma(\epsilon_s,\epsilon) &=&
	\sigma(g\epsilon_s,g\epsilon),
	\label{gtrianglesigma}
\end{eqnarray}
where we recall that we only assume $g$ to be an isometry
(i.e. an arbitrary element of $O(3)$)
and not necessarily a symmetry of the sample. This property will be crucial in the following derivations. Note that Eq.~\eqref{gtrianglesigma}
has no equivalent for antiunitary elements $g$ of a 
magnetic group~\cite{Furo-25}.

\subsection{Action of a symmetry on the cross-section}
We consider now only isometries $g$ which are symmetries of the sample.
By this we mean the point group of the space group of a crystal
or the point group of a molecule. If the site of the absorbing
atom has a lower symmetry group, then this subgroup must
be used for the calculation of the spectrum, but the average
over the different sites recovers the full point 
group~\cite{BJBA}.

For example, if the system is represented by the Hamiltonian 
$H_0=\bfp^2/2m+V$, we assume that 
$(g\triangleright V)(\bfr)=V(g^{-1}\bfr)=V(\bfr)$
and the action of $g$ on the states of the system is described by 
Eq.~\eqref{gtrianglePsi}.

To compute the action of $g$ on $\sigma(\epsilon_s,\epsilon)$,
we expand the modulus square in Eq.~\eqref{scatcc2} for a given
value of $E_I$, $E_N$, $E_{N'}$ and $E_F$
(the presence of $E_{N'}$ being due to the fact that the sum over
$N$ is inside the modulus, as in Eq.~\eqref{defaimjn}), and we replace the sum over
states of energy $E$ by a projector $P(E)$ to obtain
terms of the type.
\begin{eqnarray*}
T&=&\langle I|\epsilon^*\cdot\bfr P(E_{N'}) \epsilon_s\cdot\bfr 
P(E_F)\epsilon_s^*\cdot\bfr P(E_N)\epsilon\cdot\bfr |I\rangle.
\end{eqnarray*}
For simplicity, we considered a non-degenerate initial state
$|I\rangle$. If it is degenerate, we must
average over initial states (this also is how we take 
temperature effects into account) and  $T$ becomes
\begin{eqnarray*}
\frac{1}{n_I}\Tr\Big(P(E_I)\epsilon^*\cdot\bfr P(E_{N'}) \epsilon_s\cdot\bfr 
	P(E_F)\epsilon_s^*\cdot\bfr P(E_N)\epsilon\cdot\bfr\Big),
\end{eqnarray*}
where $n_I$ is the degeneracy of the initial state.
By using the invariance of $P(E)$ under the action
of a symmetry operation $g$ (see Eq.~\eqref{gtrianglePE}),
we see that $g\triangleright T=T$, so that
\begin{eqnarray}
g\triangleright\sigma(\epsilon_s,\epsilon) &=&
\sigma(\epsilon_s,\epsilon).
	\label{Rtrianglesigma}
\end{eqnarray}
Equations~\eqref{gtrianglesigma} and \eqref{Rtrianglesigma}
lead us to the invariance
\begin{eqnarray}
	\sigma(\epsilon_s,\epsilon) &=&
	\sigma(g\epsilon_s,g\epsilon),
	\label{actionung}
\end{eqnarray}
for every symmetry operation $g$ of the sample.
These symmetry operations form a group $G$. In other words, for any isometry of the absorbing atom point group, the RIXS cross-section is the same when the incident and scattered polarizations are rotated with this same operation. 

\subsection{Symmetry reduction of the cross-section}
\label{symredsect}
To investigate the effect of symmetry on the RIXS cross-section
we rewrite its definition Eq.~\eqref{scatcc2} as:
\begin{eqnarray}
	\sigma(\epsilon_s,\epsilon) &=& \sum_{imjn} \epsilon_{si}^*\epsilon_m\epsilon_n^*\epsilon_{sj}
	a_{im,jn},
	\label{defsigmaa}
\end{eqnarray}
where $\epsilon_m$ and $\epsilon_n$ run over the three Cartesian coordinates of 
$\epsilon$ and the same for $\epsilon_s$, and 
\begin{eqnarray}
	a_{im,jn} &=& 
	\frac{r_e^2}{\hbar^2} \frac{\omega_s}{\omega}
	\sum_F\sum_{N,N'} 
	\frac{(E_I-E_N)(E_N-E_F)}{E_I-E_N+\hbar\omega + i\gamma}
	\nonumber\\&&
	\times\frac{(E_I-E_{N'})(E_{N'}-E_F)}{E_I-E_{N'}+\hbar\omega - i\gamma}
	\langle F|r_i|N\rangle
	\langle N|r_m|I\rangle
	\nonumber\\&&
	\times\langle I|r_n|N'\rangle
\underline{}	\langle N'|r_j|F\rangle
	\delta_E,
	\label{defaimjn}
\end{eqnarray}
where $r_i$, $r_m$, $r_j$ and $r_n$ run over the Cartesian coordinates of $\bfr$.
From this definition we see that $a_{im,jn}$ is Hermitian
(i.e. $a_{im,jn}^*=a_{jn,im}$), so that $a_{im,jn}$
has a priori 81 components.

In the presence of non-trivial symmetry, relations
exist between different $a_{im,jn}$ and we call
\emph{fundamental spectra}
a set of real or imaginary parts of some $a_{im,jn}$ from which all
non-zero $a_{im,jn}$ (and thus all RIXS spectra) can
be computed. This is similar to the fundamental spectra defined
by Thole and Van der Laan for photoemission~\cite{Thole-91}. A specific $a_{im,jn}$
will provide one fundamental spectrum if it is real or purely imaginary
and two fundamental spectra if it has non-zero real and imaginary parts.

We shall see in section~\ref{polidsect}
that the invariance property of Eq.~\eqref{actionung}
implies 
\begin{eqnarray}
	a_{im,jn} = \sum_{i'm'j'n'}N_{imjn,i'm'j'n'} a_{i'm',j'n'},
	\label{invariancea}
\end{eqnarray}
where
\begin{eqnarray}
N_{imjn,i'm'j'n'} &=& \frac{1}{|G|}
	\sum_{g\in G} 
g_{ii'}g_{mm'}g_{jj'}g_{nn'}.
	\label{invariancemat}
\end{eqnarray}

In section~\ref{polidsect} we prove that, thanks to the average of $G$ in
Eq.~\eqref{invariancemat}, the left hand side of Eq.~\eqref{invariancea}
has all the symmetries of the group $G$ even if $ a_{i'm',j'n'}$
has no symmetry. In other words, the invariance under $G$ is
entirely due to the properties of $N_{imjn,i'm'j'n'}$.

We give now the practical method to determine the symmetries
of $a_{im,jn}$, which can easily be implemented with
standard symbolic computing packages.
\begin{itemize}
	\item Compute $N_{imjn,i'm'j'n'}$ by Eq.~\eqref{invariancemat}.
	\item Transform $N_{imjn,i'm'j'n'}$
	into a $81\times 81$ matrix denoted by $N^G$. For this,
	associate a number $k$ between 1 and 81 to each quadruple
	$(i,m,j,n)$. For example $k=n+3(j-1)+9(m-1)+27(i-1)$ and
	define $N^G_{kk'}=N_{imjn,i'm'j'n'}$.
	\item Compute the number $n_G$ of 
	fundamental spectra of $a_{im,jn}$, which is
	the rank of $N^G$ and also the trace of $N^G$ since
	$N^G$ is a projection (i.e. $(N^G)^2=N^G$ and $(N^G)^\dagger=N^G$).
	\item Gather in the rectangular matrix $L^G$
	the different non-zero lines of $N^G$ (i.e. if the same
	line appears several times in $N^G$ it appears only
	once in $L^G$). Let $n_L$ be the number of lines of $L^G$
	(note that $n_L\ge n_G$).
	\item Compute the null space of the transpose of $L^G$, 
	which gives you $n_L-n_G$ relations between lines by which
	you can write these lines in terms of $n_G$ independent ones.
	\item For each line $L$ of $L^G$, identify the index $k$ of the lines of
	$N^G$ which are equal to $L$ and transform back $k$
	into $(i,m,j,n)$ to express this equality of lines as an equality
	between some $a_{im,jn}$.
\end{itemize}
For the example of $D_{6h}$, we shall see that
the rank of $N^G$ is $n_G=10$,
the number of its non-zero lines is 21,
the number of different non-zero lines of $N^G$ is $n_L=11$
and there is one relation between lines.

We can also determine the numbers $n_\mathrm{Re}$ and $n_\mathrm{Im}$ 
of non-zero real and imaginary fundamental spectra by
noticing that the matrix $P$ representing the permutation
$(imjn)\to(jnim)$ commutes with $N$ and satisfies $P^2=1$.
From $P a_{im,jn}=a_{jn,im}=a_{im,jn}^*$ we see that
the number of independent real and imaginary parts in the tensor
 $a_{im,jn}$ are
$n_\mathrm{Re}=\Tr\big((1+P)N/2\big)$ and
$n_\mathrm{Im}=\Tr\big((1-P)N/2\big)$, respectively,
with $n_\mathrm{Re}+n_\mathrm{Im}=n_G$.

Table~\ref{crystptgroup} gives for each crystallographic point group, the
number of matrix elements $a_{im,jn}$ which are not zero by symmetry,
the number $n_L$ of matrix elements which are different 
(in the sense that any matrix element is equal to one of those)
and the number $n_G$ of fundamental spectra (in the sense that
any matrix element can be written as a linear combination of those).
The symbol $\simeq$ in Table~\ref{crystptgroup} means that,
if the standard reference frame~\cite{Altmann} is used (in particular 
the high symmetry axis is along the $z$ axis), then all the
matrix elements $a_{im,jn}$ of these two groups enjoy the same relations. 
The only difficulty comes from the case of $C_{3v}$ which
has two usual reference frames~\cite{Altmann}: the $1A$ parameters or
the $1B$ parameters. With $1B$ parameters (where the mirror plane
is perpendicular to the $x$ axis), the
matrix elements $a_{im,jn}$ have the same relations as those of $C_{3v}$ and $D_3$
(see Section~\ref{D3sect}).
With $1A$ parameters (where the mirror plane
is perpendicular to the $y$ axis), the matrix elements have different 
relations (see Section~\ref{C3vAsect}).
	\begin{table}
	\caption{Number of possibly non-zero matrix elements
		(i.e. that do not vanish by symmetry),
		number $n_L$ of different matrix elements,
		number $n_G$ of fundamental spectra and number $n_\mathrm{Im}$
		of imaginary components
		for all crystallographic point groups,
		$SO(3)$ and $O(3)$. For an isotropic sample, $n_S$ is the number of 
		spectra needed to calculate the full angular dependence
		and $n_C$ is the number of spectra needed to calculate the
		angular dependence when the polarization of the scattered beam
		is not measured, as explained in Section~\ref{sympowsect}. The meaning of $\simeq$
		is explained at the end of Section~\ref{symredsect}.
	}
\label{crystptgroup}
\begin{center}
	\bgroup
	\def\arraystretch{1.3}
	\begin{tabular}{|c|c|c|c|c|c|c|} \hline
		Point groups & $\not=0$ & $n_L$ & $n_G$ & $n_\mathrm{Im}$ & $n_S$ & $n_C$\\
		\hline
		$C_1$$\simeq$$C_i$ & 81 & 81 & 81 & 36 & 21 & 15 \\
		\hline
		$C_2$$\simeq$$C_s$$\simeq$$C_{2h}$ & 41 & 41 & 41 & 16 & 17 & 13 \\
		\hline
		$D_2$$\simeq$$C_{2v}$$\simeq$$D_{2h}$  & 21 & 21 & 21 & 6 & 15 & 12 \\
		\hline
		$C_4$$\simeq$$S_4$$\simeq$$C_{4h}$ & 41 & 31 & 21 & 8 & 10 & 7 \\
		\hline
		$D_4$$\simeq$$C_{4v}$$\simeq$$D_{2d}$$\simeq$$D_{4h}$ & 21 & 11 & 11 & 2 & 9 & 7 \\
		\hline
		$C_3$$\simeq$$S_6$  & 73 & 47 & 27 & 12 & 11 & 7 \\
		\hline
		$D_3$$\simeq$$D_{3d}$, $C_{3v}$  & 37 & 19 & 14 & 4 & 8 & 6 \\
		\hline
		$C_6$$\simeq$$C_{3h}$$\simeq$$C_{6h}$ & 41 & 31 & 19 & 8 & 9 & 6 \\
		\hline
		$D_6$$\simeq$$C_{6v}$$\simeq$$D_{3h}$$\simeq$$D_{6h}$ & 21 & 11 & 10 & 2 & 8 & 6 \\
		\hline
		$T$$\simeq$$T_h$ & 21 & 7 & 7 & 2 & 5 & 4 \\
		\hline
		$O$$\simeq$$T_d$$\simeq$$O_h$ & 21 & 4 & 4 & 0 & 4 & 3 \\
		\hline
		$SO(3)$$\simeq$$O(3)$ & 21 & 4 & 3 & 0 & 3 & 2 \\
		\hline
	\end{tabular}
	\egroup
	\vskip 3mm
\end{center}
\end{table}

Several other examples will be given below.

\section{Alternative way to the fundamental spectra}
\label{Alternative-sect}

In this section we show that the number of fundamental spectra
can be understood and obtained by hand using spherical tensors.
We first define spherical parameters and then
we use them to calculate circular dichroism and to 
compute the average cross-section over equivalent
sites in a crystal or a molecule.

\subsection{Definition of spherical parameters}

Instead of fundamental spectra, we can define
the \emph{spherical parameters} $S^{abc}_\gamma$, which 
were defined in~\cite{Juhin-14}
by
\begin{eqnarray}
	S^{abc}_\gamma &=& 
	\sum_{\alpha,\beta}\sum_{\mu\nu\mu'\nu'} (1\mu'1\nu|a\alpha)(1\mu1\nu'|b\beta)
	(a\alpha b\beta|c\gamma)
    \nonumber\\&&\times A_{i\mu'} A_{m\nu} A_{n\mu}A_{j\nu'} \,a_{im,jn},
    \label{Sabcdef}
\end{eqnarray}
(corresponding to $S^{g_1g_2a}_L$ in Ref.~\onlinecite{Juhin-14}),
where $a$ and $b$ take the values 0, 1, 2 and
$c$ runs from $|a-b|$ to $a+b$, so that 
$c$ takes values from 0 to 4.

For the example of a sample with symmetry $O$, $T_d$ or $O_h$,
the non-zero spherical parameters of this approach are
\begin{eqnarray*}
	S^{000}_0 &=& a_{11,11} + 2 a_{11,22},\\
	S^{110}_0 &=& -\sqrt{3} (a_{12,12} - a_{12,21}),\\
	S^{220}_0 &=& \frac{1}{\sqrt{5}} 
    (2a_{11,11}-2a_{11,22}+3a_{12,12} + 3 a_{12,21}),\\
	S^{224}_4 &=& S^{224}_{-4} = \sqrt{\frac{5}{14}} \,S^{224}_0
    \\&=&
    \frac{1}{2} 
    (a_{11,11}-a_{11,22}-a_{12,12} - a_{12,21}).
\end{eqnarray*}
The relation between the components of the
spherical tensor $S^{224}$ is a well-known
consequence of cubic symmetry~\cite{BrouderAXAS}.
It illustrates the fact that several spherical
parameters $S^{abc}_\gamma$ can be proportional.
Of course, the number of independent spherical
parameters is the same as the number of fundamental spectra.

The relation $a_{jn,im}=a_{im,jn}^*$ implies
\begin{eqnarray*}
	(S^{abc}_\gamma)^* &=& (-1)^{a+b+c+\gamma} S^{abc}_{-\gamma},
\end{eqnarray*}
which shows that $S^{abc}_0$ is real if $a+b+c$ is even and
purely imaginary if $a+b+c$ is odd.
 
To describe the action of an isometry $g$ on the spherical parameters
$S^{abc}_\gamma$, we first note that any isometry can be written
$g=(-1)^{n_g} R$, where $R$ is a rotation matrix and
$n_g=0$ or $1$ if $\det g=1$ ($g$ is said to be even)
or $-1$ ($g$ is said to be odd), respectively.  
The rotation $R$ will be called the \emph{rotation part} of $g$.

The fact that the columns of matrix $A$ can be considered as
polarization vectors leads to the group action
\begin{eqnarray*}
	g^{-1}\triangleright S^{abc}_\gamma &=& 
   \sum_{\gamma'=-c}^c S^{abc}_{\gamma'} D^c_{\gamma'\gamma}(R),
\end{eqnarray*}
where $R$ is the rotation part of $g$ and where
$D^c_{\gamma'\gamma}(R)$ is the Wigner rotation matrix
whose arguments are the same Euler angles as those of $R$.
The part $(-1)^{n_g}$ of $g$ disappears because the group
action involves a product of four $g$.

When the symmetry group $G$ is only made of even 
isometries, building group invariants from
$a_{im,jn}$ reduces to building group invariants
from spherical harmonics $Y_c^\gamma$ and the
number $n_G(c)$ of such invariants is given
by the number of fully symmetric representations
(denoted $A_{1g}$ or $A_1$ or $A$)
of $G$ indicated in the section ``Subduction from $O(3)$'' of
 Altmann and Herzig's tables~\cite{Altmann}.
 These group invariants can also be obtained
 explicitly by considering the group averages
 \begin{eqnarray*}
\langle Y_c^\gamma\rangle &=& \frac{1}{|G|}\sum_{g\in G} 
   Y_c^{\gamma'} D^c_{\gamma'\gamma}(R).
\end{eqnarray*}
These averages for $c=2$ and $4$ were constructed in Ref.~\cite{BrouderAXAS}.

The spherical tensor decomposition described by
Eq.~\eqref{Sabcdef} can be summarized by 
\begin{eqnarray}
	\mathbf{1}^{\otimes 4} &=& (\mathbf{1}^{\otimes 2} )^{\otimes 2}= 
	(\mathbf{0}\oplus\mathbf{1}\oplus\mathbf{2})^{\otimes 2}
	\nonumber\\&=&
	(3{\times}\mathbf{0})\oplus (6{\times}\mathbf{1})\oplus (6{\times}\mathbf{2})
	\oplus (3{\times}\mathbf{3})\oplus\mathbf{4},
	\label{tensordecomposition}
\end{eqnarray}
which means that $a_{im,jn}$ can be seen as the sum of
three spherically symmetric tensors, six 
tensors of rank 1, six tensors of rank 2,
 three tensors of rank 3 and one tensor of rank 4.
 The total number of fundamental spectra is then
 the weighted sum of the numbers $n_G(c)$:
\begin{eqnarray*}
n_G = 3n_G(0)+6n_G(1)+6n_G(2)+3n_G(3)+n_G(4).
\end{eqnarray*}

The procedure must be modified when some elements $g$ of $G$
are odd, which is the case of the groups that
are not in the first position in a line of 
Table~\ref{crystptgroup}. Indeed, Altmann and Herzig's tables
are constructed such that an inversion cancels
all spherical tensors of odd rank, whereas it has 
no effect on the RIXS tensor. 
Therefore, these tables must be used with a group $G'$
derived from $G$ as follows.
Since $G$ contains an odd element, $G$ can be written
as the union of a set $G^+$ of even elements
and a set $G^-$ of odd elements. 
Since the inversion $I$ commutes with all rotations
and satisfies $I^2=1$, we see that $G^+$ is a 
subgroup of $G$ and has the same number of elements as $G^-$.
If $I$ is an element of $G^-$, then $IG^-=G^+$,
where $IG^-$ is the set of $-g_{ij}$ with $g\in G^-$. In that case
we take $G'=G^+$ (because the average Eq.~\eqref{invariancemat}
 over $G$ is the same as the average over $G^+$).
If $I$ is not an element of $G^-$, then
$G'=G^+\cup IG^-$ is a group without odd elements,
which has the same number of elements as $G$ and 
such that $G^+$ is a subgroup of $G'$.
The number $n_G$ of fundamental spectra of $G$ 
is then the weighted sum of $n_{G'}(c)$.
In Table~\ref{crystptgroup} the first group of each line
is the $G'$ for all groups of that line.

Conversely, if the spherical parameters are constructed as
linear combinations of $Y^\gamma_c$ for all $c$ such that $n_{G'}(c)\not=0$,
they can be transformed into relations for $a_{im,jn}$
by inverting Eq.~\eqref{Sabcdef}:
\begin{eqnarray}
a_{im,jn} &=& 
	\sum_{a\alpha b\beta c\gamma}\sum_{\mu\nu\mu'\nu'} (1\mu'1\nu|a\alpha)(1\mu1\nu'|b\beta)(a\alpha b\beta|c\gamma)\nonumber\\&&\times 
	A_{i\mu'}^* A_{m\nu}^* A_{n\mu}^*A_{j\nu'}^* S^{abc}_\gamma .
    \label{adeSabc}
\end{eqnarray} 

\subsection{Circular dichroism and permutation symmetry}
\label{Ydefsect}
Circular dichroism is the difference between the
spectra obtained from left- and right-polarized incident beams.
This is now a well-established
technique~\cite{Haraki-25,Nag-25,Takegami-25,Furo-25}.

Circular dichroism in RIXS measures the difference 
\begin{eqnarray}
	\Delta\sigma(\epsilon_s,\epsilon) &=&
	\sigma(\epsilon_s,\epsilon)-\sigma(\epsilon_s,\epsilon^*).
\end{eqnarray}
By Eq.~\eqref{defsigmaa} we see that 
\begin{eqnarray*}
	\Delta\sigma(\epsilon_s,\epsilon) &=& 
    \sum_{imjn} \epsilon_{si}^*\epsilon_m\epsilon_n^*\epsilon_{sj}
	(a_{im,jn}-a_{in,jm}).
\end{eqnarray*}
Circular dichroism is 
usually normalized with respect to the average
of $\sigma(\epsilon_s,\epsilon)$ and $\sigma(\epsilon_s,\epsilon^*)$
which is obtained by $(a_{im,jn}+a_{in,jm})/2$.
The basic remark of this section is to notice that
$a_{in,jm}$ is obtained from $a_{im,jn}$
by the permutation $(imjn)\to(injm)$.

We saw by Eq.~\eqref{adeSabc} that 
\begin{eqnarray*}
	a_{im,jn} &=& \sum_{abc\gamma} Y_{imjn}^{ab}(c,\gamma)S^{abc}_\gamma,
\end{eqnarray*} 
where
\begin{eqnarray*}
	Y_{imjn}^{ab}(c,\gamma) &=& 
	\sum_{\alpha \beta}\sum_{\mu\nu\mu'\nu'} (1\mu'1\nu|a\alpha)(1\mu1\nu'|b\beta)(a\alpha b\beta|c\gamma)\\&&\times 
	A_{i\mu'}^* A_{m\nu}^* A_{n\mu}^*A_{j\nu'}^* .
\end{eqnarray*} 
The determination of circular dichroism in RIXS by $a_{im,jn}-a_{in,jm}$
amounts to the calculation of 
$Y_{imjn}^{ab}(c,\gamma)-Y_{injm}^{ab}(c,\gamma)$.
The quantity $Y_{injm}^{ab}(c,\gamma)$ is obtained from $Y_{imjn}^{ab}(c,\gamma)$
by the permutation $(imjn)\to(injm)$. 
Spherical tensor recoupling techniques allow us to write
this permutation in terms of $9j$-symbols~\cite[p.~70]{VMK}
\begin{eqnarray*}
	Y^{ab}_{injm}(c,\gamma) &=& 
 	\sum_{d,e}\Pi_{abde}\ninej{1}{1}{a}{1}{1}{b}{d}{e}{c}
	Y^{de}_{imjn}(c,\gamma),
\end{eqnarray*}
where 
\begin{eqnarray*}
\Pi_{abde} &=& \sqrt{(2a+1)(2b+1)(2d+1)(2e+1)}.
\end{eqnarray*}
To describe the sum over $d,e$, recall that
Eq.~\eqref{tensordecomposition} describes
the multiplicity of rank $c$ in the tensor decomposition of $\bfun^{\otimes 4}$.
This multiplicity
is $T(c)=3,6,6,3,1$ for $c=0,1,2,3,4$, respectively,
where $T(c)$ is the number of pairs $(a,b)$, 
 $a$ and $b$ taking values in $\{0,1,2\}$, 
such that the triangle rule $|a-b|\le c \le a+b$ is satisfied. 
For example, for $c=2$, the 6 pairs are
$(a,b)=(0,2), (1,1), (1,2), (2,0), (2,1), (2,2)$.
The sum over $d,e$ is the sum over
these $T(c)$ pairs.

Other permutations have a physical meaning:
the measurement of the circular dichroism of the scattered beam
is obtained by the permutation $(imjn)\to(jmin)$,
the interchange of the incident and scattered beam 
is described by the permutation $(imjn)\to(njmi)$,
which can be obtained by combining the permutations
\begin{eqnarray*}
	Y^{ab}_{jnim}(c,\gamma) &=& 
	(-1)^cY^{ab}_{imjn}(c,\gamma),\\
	Y^{ab}_{mijn}(c,\gamma) &=& 
(-1)^aY^{ab}_{imjn}(c,\gamma),\\
	Y^{ab}_{imnj}(c,\gamma) &=& 
(-1)^bY^{ab}_{imjn}(c,\gamma).
\end{eqnarray*}

In conclusion, every permutation of $(imjn)$
in $Y^{ab}_{imjn}(c,\gamma)$ can be expressed as a linear
combination of $Y^{de}_{imjn}(c,\gamma)$. In group theoretical parlance,
for each $c=0,\dots,4$,
the spherical tensors $Y^{ab}_{imjn}(c,\gamma)$ form a $T(c)$-dimensional representation
of the symmetric group $S_4$ of permutations of four elements.
The decomposition of these $T(c)$ tensors into 
irreducible representations of $S_4$ allows us to pick up
the permutation symmetry required for 
the physical quantity of interest
(circular dichroism of the incident or scattered beam,
average spectrum of the incident or scattered beam, etc.).

Recall that $S_4$ is a group of order 24 with 5 irreducible
representations: two of them are one dimensional (the trivial representation
denoted by $(4)$ and the sign representation denoted by (1,1,1,1)),
one is two-dimensional (denoted by $(2,2)$) and
two are three-dimensional (the standard representation
denoted by $(3,1)$ and the product of the sign and standard representations
denoted by $(2,1,1)$).
An explicit calculation gives us the decomposition of 
$T(c)$-dimensional representation generated by $Y^{ab}(c)$
into irreducible representations of $S_4$:
\begin{eqnarray*}
	3{\times}\mathbf{0} &=& (4) \oplus (2,2),\\
	6{\times}\mathbf{1} &=& (3,1) \oplus (2,1,1),\\
	6{\times}\mathbf{2} &=& (4)  \oplus (3,1) \oplus (2,2),\\
    3{\times}\mathbf{3} &=& (3,1),\\
    \mathbf{4} &=& (4).
\end{eqnarray*}

This combination of point group and permutation symmetries was
also found useful in the calculation of 
Coulomb matrix elements~\cite{Letouze-23}.

The conclusion of this discussion is that all linear operations that
are carried out on the 81 matrix elements of $a_{im,jn}$ can
be encapsulated into equivalent linear operations on the
(generally) much less numerous spherical parameters
$S^{abc}_\gamma$. 

A last useful relation is
\begin{eqnarray*}
	Y_{imjn}^{ab}(c,\gamma)^* &=& 
    (-1)^{c+\gamma}Y_{imjn}^{ab}(c,-\gamma).
\end{eqnarray*} 

\subsection{From site to point group symmetry}
In general, the point symmetry group $G$ of a molecule or crystal generates
several equivalent sites.
Consider a scattering atom at site $A$ with coordinates $\bfR_A$. Its local symmetry group $H$ is the
set of elements $g$ of $G$ such that $g\bfR_A=\bfR_A$.
The number of fundamental spectra for the RIXS of this atom is that of $H$.
Since our formulation is independent of the translation between atoms,
the scattering cross section of the full molecule or crystal is computed
as the average of the cross-sections of the equivalent atoms.

Here again, the use of spherical parameters will be particularly convenient.
The transition from site symmetry to point group was discussed
in detail for the case of X-ray absorption spectroscopy~\cite{BJBA}
and much of that discussion holds for RIXS. 
If $S^{abc}_\gamma(A)$ are the spherical parameters of site $A$, then
the spherical parameters $S^{abc}_\gamma(B)$ corresponding to an equivalent site $B$
obtained from $A$ by the transformation $\bfr_B=g\bfr_A$ (plus a possible translation)
where $g$ is an element of $G$ which is not in $H$, are~\cite[Eq.~(24)]{BJBA}
\begin{eqnarray*}
	S^{abc}_\gamma(B) &=& \sum_{\gamma'=-c}^c S^{abc}_{\gamma'}(A) D^c_{\gamma'\gamma}(R^{-1}),
\end{eqnarray*}
where $R^{-1}$ is the inverse of the rotation part of $g$. 
And the spherical parameters of the full molecule or crystal is
\begin{eqnarray}
	S^{abc}_\gamma &=& \frac{1}{n}\sum_{g\in K}\sum_{\gamma'=-c}^c S^{abc}_{\gamma'}(A) D^c_{\gamma'\gamma}(R^{-1}),
	\label{sommesurK}
\end{eqnarray}
where $K$ is a set of representatives of the coset of $H$ in $G$, i.e. a set of
elements of $G$ that transform $\bfr_A$ into all the sites equivalent to $A$ (up to possible translations),
and $n$ is the number of elements of $K$.

To illustrate the difference between the fundamental spectra and the spherical parameters approaches,
consider a site with symmetry $D_4$ in a cubic crystal. We choose as coset representatives
the identity, the rotation of $2\pi/3$ around the $(111)$ axis and the rotation of $2\pi/3$ around
the $(\overline{1}1\overline{1})$ axis, but the result does not depend on the representatives. 
The relation between the fundamental spectra of $O$ and $D_4$ is
\begin{eqnarray*}
	a_{11,11}^O &=& \frac23 a_{11,11}^{D_4} + \frac13 a_{33,33}^{D_4},\\
	a_{11,22}^O &=& \frac13 a_{11,22}^{D_4} + \frac23 \Re(a_{11,33}^{D_4}),\\
	a_{12,12}^O &=& \frac13 a_{12,12}^{D_4} + \frac13 a_{13,13}^{D_4} + \frac13 a_{31,31}^{D_4}  ,\\
	a_{12,21}^O &=& \frac13 a_{12,21}^{D_4} + \frac23 \Re(a_{13,31}^{D_4}).
\end{eqnarray*}
So we need to determine 9 components (among the 11 of $D_4$) to build the 4 components of $O$. 
Only the imaginary parts of the fundamental spectra of $D_4$ are not required.

The spherical parameters for $D_4$ are 3 parameters $S^{aa0}_0(D_4)$ with $a=0,1$ and $2$, 6 parameters $S^{ab2}_0(D_4)$
with $(ab)=(02), (11), (12), (20), (21), (22)$
and two parameters $S^{224}_0(D_4)$ and $S^{224}_{4}(D_4)=S^{224}_{-4}(D_4)$.

The spherical parameters of the crystal with symmetry $O$ obtained from
the average of Eq.~\eqref{sommesurK} is then the three parameters 
$S^{aa0}_0(O)=S^{aa0}_0(D_4)$ which are the
same as those of $D_4$ because a spherical tensor of rank 0 is invariant under rotation,
the six parameters $S^{ab2}_0$ vanish because no invariant of $O$ can be constructed from $c=2$
and we do not have to calculate them, 
and the non-zero spherical parameters are
\begin{eqnarray*}
	S^{224}_4(O) &=& S^{224}_{-4}(O) = \sqrt{\frac{5}{14}} \,S^{224}_0(O)
\\&=&
\frac{5}{12}S^{224}_4(D_4) + \frac{\sqrt{70}}{24}S^{224}_0(D_4).
\end{eqnarray*}
Therefore, we only have to calculate 5 spherical parameters of $D_4$
(instead of 9) to obtain the 4 spherical parameters of $O$.

\section{RIXS cross-section for various point groups}

We give now the full determination of the fundamental spectra and of the relations between the
$a_{im,jn}$ (defined in Eq.~\eqref{defaimjn}) for several symmetry groups.

\subsection{Case $C_1$ or $C_i$}
If the symmetry group is $C_1$ or $C_i$
no $a_{im,jn}$ is constrained to be zero by symmetry
and the only relation between them is $a_{jn,im}=a_{im,jn}^*$,
so that there are nine real fundamental spectra
(i.e. $a_{im,im}$) and the other fundamental spectra can be
grouped into 36 complex terms. 
The number of fundamental spectra is indeed
\begin{eqnarray*}
n_G = 3\times 1+6\times 3+6\times 5+3\times 7+9=81.
\end{eqnarray*}

\subsection{Case $C_2$, $C_s$ or $C_{2h}$}
If the symmetry group is $C_2$ or $C_s$, then
$a_{im,jn}$ is zero if exactly one or exactly three
of the indices $imjn$ are equal to 3. 
There are 16 terms where none of the indices $imjn$ are equal to 3,
24 terms where exactly two of the indices $imjn$ are equal to 3
and one term $imjn=3333$. 
The only relation between matrix elements is
$a_{jn,im}=a_{im,jn}^*$.
\begin{eqnarray*}
n_G = 3\times 1+6\times 1+6\times 3+3\times 3+5=41.
\end{eqnarray*}

\subsection{Case $D_2$ or $C_{2v}$ or $D_{2h}$}
If the symmetry group is $D_2$ or $C_{2v}$ or $D_{2h}$, the
matrix elements $a_{imjn}$ which are not zero by symmetry can 
be written as nine fundamental spectra ($a_{im,im}$ for $i$ and $m$ equal
1,2,3) that are always real
and 12 ones that can be grouped into 6 complex parameters
\begin{eqnarray*}	a_{11,22} &=& a_{22,11}^*,\,\, a_{11,33}=a_{33,11}^*,\,\, a_{12,21}=a_{21,12}^*,\\
	  a_{22,33} &=& a_{33,22}^*,\,\,  a_{23,32}=a_{32,23}^*,\,\,a_{13,31}=a_{31,13}^*.
\end{eqnarray*}
\begin{eqnarray*}
n_G = 3\times 1+6\times 1+6\times 3+3\times 3+5=41.
\end{eqnarray*}

\subsection{Case $C_4$ or $S_4$ or $C_{4h}$}
If the symmetry group is $C_4$ or $S_4$ or $C_{4h}$, 
we have seven real fundamental spectra
\begin{eqnarray*}
	a_{13,13}&=& a_{23,23},\quad a_{12,21}= a_{21,12},\\
	a_{12,12}&=& a_{21,21},\quad a_{11,22}= a_{22,11},\\
	a_{11,11}&=& a_{22,22},\quad a_{31,31}= a_{32,32},\\
	&a_{33,33},&
\end{eqnarray*}
two imaginary fundamental spectra
\begin{eqnarray*}
	a_{23,13}&=&-a_{13,23},\quad a_{32,31}=-a_{31,32},
\end{eqnarray*}
and six complex fundamental spectra
\begin{eqnarray*}
	a_{22,21}&=&-a_{11,12}=a_{21,22}^*=-a_{12,11}^*,\\
	a_{22,12}&=&-a_{11,21}=a_{12,22}^*=-a_{21,11}^*,\\
	a_{21,33}&=&-a_{12,33}=a_{33,21}^*=-a_{33,12}^*,\\
	a_{23,31}&=&-a_{13,32}=a_{31,23}^*= -a_{32,13}^*,\\
	a_{33,11}&=& a_{33,22}=a_{11,33}^*=a_{22,33}^*,\\
	a_{31,13}&=& a_{32,23}=a_{13,31}^*=a_{23,32}^*.
\end{eqnarray*}
\begin{eqnarray*}
n_G = 3\times 1+6\times 1+6\times 1+3\times 1+3=21.
\end{eqnarray*}

\subsection{Case $D_4$ or $C_{4v}$ or $D_{2d}$ or  $D_{4h}$}
\label{D4sect}
If the symmetry group is $D_4$ or $C_{4v}$ or $D_{2d}$ or  $D_{4h}$, 
there are seven real fundamental spectra
\begin{eqnarray*}
	a_{13,13} &=& a_{23,23},\quad a_{31,31} = a_{32,32},\quad a_{12,21} = a_{21,12},\\
	a_{12,12} &=& a_{21,21},\quad a_{11,22} = a_{22,11}, \quad a_{11,11} = a_{22,22}, \\
    a_{33,33},
\end{eqnarray*}
and four complex parameters corresponding to eight fundamental spectra
\begin{eqnarray*}
	a_{13,31} &=& a_{23,32} = a_{31,13}^*=a_{32,23}^*\\
	a_{11,33} &=& a_{22,33} = a_{33,22}^*=a_{33,11}^*.\\
\end{eqnarray*}
\begin{eqnarray*}
n_G = 3\times 1+6\times 0+6\times 1+3\times 0+2=11.
\end{eqnarray*}

\subsection{Case $D_3$ or $D_{3d}$ or $C_{3v}$ with $1B$ reference frame}
\label{D3sect}
If the symmetry group is $D_3$ or $D_{3d}$ or $C_{3v}$ with $1B$ reference frame, the fundamental spectra are six real parameters
\begin{eqnarray*}
	a_{13,13} &=& a_{23,23},\quad a_{31,31} = a_{32,32},\quad a_{12,21} = a_{21,12},\\
	a_{12,12} &=& a_{21,21},\quad a_{11,22} = a_{22,11},\quad a_{33,33},
\end{eqnarray*}
and four complex parameters
\begin{eqnarray*}
	a_{11,32} &=& a_{12,31} = a_{21,31}=-a_{22,32}\\
    &=& a_{31,12}^* = a_{31,21}^*=a_{32,11}^*=-a_{32,22}^*,\\
	a_{11,23} &=& a_{12,13} = a_{21,13}=-a_{22,23}\\
	&=& a_{13,12}^* = a_{13,21}^*=a_{23,11}^*=-a_{23,22}^*,\\
	a_{11,33} &=& a_{22,33}=a_{33,11}^* = a_{33,22}^*,\\
	a_{13,31} &=& a_{23,32}=a_{31,13}^* = a_{32,23}^*,
\end{eqnarray*}
with the remaining non-zero matrix elements
\begin{eqnarray*}
	a_{11,11} &=& a_{22,22} = a_{11,22}+a_{12,12}+a_{12,21}.
\end{eqnarray*}
\begin{eqnarray*}
n_G = 3\times 1+6\times 0+6\times 1+3\times 1+2=14.
\end{eqnarray*}

\subsection{Case $C_{3v}$ with $1A$ reference frame}
\label{C3vAsect}
If the symmetry group is $C_{3v}$ with $1A$ reference frame, the fundamental spectra are six real parameters
\begin{eqnarray*}
	a_{13,13} &=& a_{23,23}\quad a_{31,31} = a_{32,32}\quad a_{12,21} = a_{21,12},\\
	a_{12,12} &=& a_{21,21},\quad a_{11,22} = a_{22,11},\quad a_{33,33},
\end{eqnarray*}
and four complex parameters
\begin{eqnarray*}
	a_{12,23} &=& a_{21,23}=a_{22,13} =-a_{11,13}\\
	&=& a_{13,22}^* = a_{23,12}^*=a_{23,21}^*=-a_{13,11}^*,\\	
	a_{12,32} &=& a_{21,32}=a_{22,31}= -a_{11,31}\\	
	&=& a_{31,22}^* = a_{32,12}^*=a_{32,21}^*=-a_{31,11}^*,\\	
	a_{33,11} &=& a_{33,22}=a_{11,33}^* = a_{22,33}^*,\\
	a_{31,13} &=& a_{32,23}=a_{13,31}^* = a_{23,32}^*,
\end{eqnarray*}
with the remaining non-zero matrix elements
\begin{eqnarray*}
	a_{11,11} &=& a_{22,22} = a_{11,22}+a_{12,12}+a_{12,21}.
\end{eqnarray*}
\begin{eqnarray*}
n_G = 3\times 1+6\times 0+6\times 1+3\times 1+2=14.
\end{eqnarray*}

\subsection{$a_{imjn}$ for $D_6$, $C_{6v}$, $D_{3h}$ and $D_{6h}$}
If the symmetry group is $D_6$, $C_{6v}$, $D_{3h}$ or $D_{6h}$, we see from the decomposition of 
Eq.~\eqref{tensordecomposition} that we have ten fundamental spectra
(three coming from tensors of rank 0, six from tensors of rank 2
and 1 from the tensor of rank 4).
The first six real ones are:
\begin{eqnarray*}
	a_{11,22} &=& a_{22,11},\quad
	a_{12,12} = a_{21,21},\quad
	a_{12,21} = a_{21,12},\\
	a_{13,13} &=& a_{23,23},\quad
	a_{31,31}= a_{32,32},\quad a_{33,33},\\
\end{eqnarray*}
The other four are grouped into two complex parameters
\begin{eqnarray*}
	a_{11,33} &=& a_{22,33}=a_{33,11}^* = a_{33,22}^*,\\
	a_{13,31} &=& a_{23,32}=a_{31,13}^* = a_{32,23}^*.
\end{eqnarray*}
The remaining non-zero matrix elements $a_{im,jn}$ are
\begin{eqnarray*}
	a_{11,11} &=& a_{22,22}=a_{11,22}+a_{12,12}+a_{12,21}.
\end{eqnarray*}
\begin{eqnarray*}
n_G = 3\times 1+6\times 0+6\times 1+3\times 0+1=10.
\end{eqnarray*}

\subsection{$a_{imjn}$ for $T$ or $T_h$}
If the symmetry group $G$ is $T$ or $T_h$, the explicit calculation of the matrix $N^G_{kk'}$
shows that there are 21 non-zero $a_{im,jn}$ that can be expressed in terms of 7
fundamental spectra. Three of them are real
\begin{eqnarray*}
	a_{11,11} &=& a_{22,22} = a_{33,33},\\  
	a_{12,12} &=& a_{23,23} = a_{31,31},\\
	a_{13,13} &=& a_{21,21} = a_{32,32},\\
\end{eqnarray*}
and the remaining four can be grouped into two complex numbers:
\begin{eqnarray*}
	a_{11,22} &=& a_{22,33} = a_{33,11} = a_{11,33}^* = a_{22,11}^* = a_{33,22}^*,\\
	a_{12,21} &=& a_{23,32} = a_{31,13} = a_{13,31}^* = a_{21,12}^* = a_{32,23}^*.  
\end{eqnarray*}
\begin{eqnarray*}
n_G = 3\times 1+6\times 0+6\times 0+3\times 1+1=7.
\end{eqnarray*}

\subsection{$a_{imjn}$ for $O_h$ or $O$ or $T_d$}
If the symmetry group $G$ is $O_h$ or $O$ or $T_d$, the explicit calculation of the matrix $N^G_{kk'}$
shows that there are 21 non-zero $a_{im,jn}$ that can be expressed in terms of 4
real fundamental spectra:
\begin{eqnarray*}
	a_{11,11} &=& a_{22,22} = a_{33,33},\\  
	a_{11,22} &=& a_{11,33} = a_{22,33} = a_{22,11} = a_{33,11} = a_{33,22},\\
	a_{12,12} &=& a_{13,13} = a_{23,23} = a_{21,21} = a_{31,31} = a_{32,32},\\
	a_{12,21} &=& a_{13,31} = a_{23,32} = a_{21,12} = a_{31,13} = a_{32,23}.
\end{eqnarray*}
They are real because, for instance, 
$a_{im,jn}^*=a_{jn,im}$ implies 
$a_{12,21}^*=a_{21,12}=a_{12,21}$.
\begin{eqnarray*}
n_G = 3\times 1+6\times 0+6\times 0+3\times 0+1=4.
\end{eqnarray*}

An alternative mathematical approach to this question was proposed
by Perla Azzi and coll.~\cite{Azzi-24}
for the elastic tensor, which is also a fourth-rank
tensor, but with a different symmetry with respect
to the interchange of indices. They also investigated the distance
with respect to cubic symmetry.

\subsection{$a_{imjn}$ for spherical samples}
\label{sphersect}
For a spherical sample (e.g. an atom), we cannot use 
directly the average given by Eq.~\eqref{invariancemat}
because the number $|G|$ of rotations in infinite. 
However, since $SO(3)$ is a compact group, 
the average is obtained as a normalized integral over Euler angles
 $\alpha$, $\beta$ and $\gamma$~\cite{Serre-group}:
\begin{eqnarray}
	N_{imjn,i'm'j'n'} &=& \langle
	g_{ii'}g_{mm'}g_{jj'}g_{nn'}\rangle.
	\label{invariancematsphe}
\end{eqnarray}
To compute this spherical average, we
write each rotation matrix $g$ in terms of the 
Wigner rotation matrix $D^1_{\mu\mu'}$, for which averages over
angles are well known. 

The relation giving the Cartesian basis $(e_x,e_y,e_z)$ in terms
of the spherical basis $(e_{-1},e_0,e_{+1})$ is 
\begin{eqnarray}
	\left( \begin{array}{c}
		e_x \\
		e_y \\
		e_z
	\end{array}\right)
	&=& \left( \begin{array}{ccc}
		\frac{1}{\sqrt2} & 0 & -\frac{1}{\sqrt2}\\
		\frac{i}{\sqrt2} & 0 & \frac{i}{\sqrt2}\\
		0 & 1 & 0
	\end{array}\right)
	\left( \begin{array}{c}
		e_{-1} \\
		e_{0} \\
		e_{+1}
	\end{array}\right),
	\label{defmatriceA}
\end{eqnarray}
where the $3\times 3$ matrix will be denoted by $A$. 
Thus, the rotation matrix $g$ is related to the
Wigner rotation matrix $D^1$ by 
\begin{eqnarray}
	g_{ij}=\sum_{\mu\mu'} A_{i\mu} A_{j\mu'}^* D^1_{\mu\mu'}.
	\label{gWigner}
\end{eqnarray}
This way, the spherical average of Eq.~\eqref{invariancematsphe}
is reduced to the angular average of a product of four Wigner matrices
(implicitly) expressed
in terms of Euler angles:
\begin{eqnarray*}
	N'	&=&
	\langle D^1_{\mu_1,\mu'_1} D^1_{\mu_2,\mu'_2} D^1_{\mu_3,\mu'_3} D^1_{\mu_4,\mu'_4} \rangle
	\\	&=&
	\frac{1}{8\pi^2} \int_0^{2\pi}d\alpha \int_0^\pi \sin\beta d\beta
	\int_0^{2\pi}d\gamma 
	D^1_{\mu_1,\mu'_1}
	\\&&\times D^1_{\mu_2,\mu'_2}
	D^1_{\mu_3,\mu'_3} D^1_{\mu_4,\mu'_4}.
\end{eqnarray*}
To compute this, we write the product of two Wigner matrices
as a sum of Wigner matrices~\cite[p.~84]{VMK} weighted by
Clebsch-Gordan coefficients,
then we use the angular average of a product of two Wigner matrices~\cite[p.~96]{VMK}
and obtain
\begin{eqnarray*}
	N'	&=&
	\sum_{b=0}^2\frac{(-1)^{\beta'-\beta}}{2b+1}
	(1\mu_1,1\mu_2|b\beta)(1\mu'_1,1\mu'_2|b\beta') 
	\\&&\times 
	(1\mu_3,1\mu_4|b-\beta)(1\mu'_3,1\mu'_4|b-\beta'),
\end{eqnarray*}
where $\beta=\mu_1+\mu_2=-\mu_3-\mu_4$ and similarly for $\beta'$.
By multiplying with $A$ and $A^*$ and summing over $\mu_j$ we find
21 non-zero $a_{im,jn}$ that can be expressed in terms of 3
fundamental spectra 
\begin{eqnarray*}
	a_{11,22} &=& a_{11,33} = a_{22,33} = a_{22,11} = a_{33,11} = a_{33,22},\\
	a_{12,12} &=& a_{13,13} = a_{23,23} = a_{21,21} = a_{31,31} = a_{32,32},\\
	a_{12,21} &=& a_{13,31} = a_{23,32} = a_{21,12} = a_{31,13} = a_{32,23}.  
\end{eqnarray*}
The remaining non-zero matrix elements are
\begin{eqnarray*}
	a_{11,11} = a_{22,22}= a_{33,33}=a_{11,22}+a_{12,12}+a_{12,21}.
\end{eqnarray*}
If we compare with a cubic sample, we see
that the first fundamental spectrum of a cubic sample
becomes the sum of the three other ones 
in the spherical case.

With these parameters, we recover the fact~\cite{Juhin-14} that angular dependence of
a RIXS spectrum of a powder is entirely determined by coefficients
of $|\epsilon\cdot\epsilon_s|^2$ and $|\epsilon\cdot\epsilon_s^*|^2$.

If we put these coefficients $a_{im,jn}$ in the 
RIXS cross-section of Eq.~\eqref{defsigmaa}, we find that
for general (i.e. possibly non normalized) $\epsilon$ and $\epsilon_s$
the RIXS cross-section of a powder is
\begin{eqnarray}
	\sigma(\epsilon_s,\epsilon) &=& a_{11,22}|\epsilon\cdot\epsilon_s^*|^2
	+ a_{12,21}|\epsilon\cdot\epsilon_s|^2
	\nonumber\\&&
	+a_{12,12} ||\epsilon||^2||\epsilon_s||^2.
	\label{generalpowder}
\end{eqnarray}
This is a linear combination of the three possible spherically symmetric
polynomials of degree four that can be constructed from 
 two complex vectors $\epsilon$ and $\epsilon_s$.
For normalized polarization vectors,
we recover the fact~\cite{Juhin-14} that the angular dependence of
a RIXS spectrum of a powder is determined by the coefficients
of $|\epsilon\cdot\epsilon_s|^2$ and $|\epsilon\cdot\epsilon_s^*|^2$.

If the polarization of the scattered beam is not measured,
we can average over its normalized polarizations perpendicular
to the unit vector $\widehat\bfk_s$ along the scattered beam~\cite{Juhin-14} to obtain
\begin{eqnarray}
	\sigma(\widehat\bfk_s,\epsilon) &=& 
	\frac12 (a_{11,22} + 2 a_{12,12} + a_{12,21})||\epsilon||^2
	\nonumber\\&&
	-\frac12 (a_{11,22} + a_{12,21})|\epsilon\cdot\widehat\bfk_s|^2.
	\label{unpolarizedpowder}
\end{eqnarray}

\section{RIXS measurement of a powder}\label{sec:rixs-powder}
In this section we deal in detail with the way RIXS measurements
of an isotropic sample
depend on the polarizations (and indirectly on the directions)
of the incident and scattered beams.

For X-ray absorption spectra, it is well known that the spectrum $\langle \sigma_{\mathrm{XAS}}\rangle_{\mathrm{powder}}$
of a powder does not depend on polarization and can be calculated as the average of three spectra
$\sigma_{\mathrm{XAS}}(\epsilon)$ calculated for a single crystal 
with a polarization vector $\epsilon$ along $x$, $y$ and $z$~\cite{Holmgaard-17}:
\begin{eqnarray*}
	\langle \sigma_{\mathrm{XAS}}\rangle_{\mathrm{powder}} &=& 
    \frac{\sigma_{\mathrm{XAS}}(\epsilon_x)+\sigma_{\mathrm{XAS}}(\epsilon_y)
    +\sigma_{\mathrm{XAS}}(\epsilon_z)}{3}.
\end{eqnarray*}
Moreover, if the sample has cubic symmetry, then all three spectra are identical
and only one spectrum must be computed to obtain
$\langle \sigma_{\mathrm{XAS}}\rangle_{\mathrm{powder}}=\sigma_{\mathrm{XAS}}(\epsilon_x)$.

For a magnetic powder the calculation of an X-ray absorption spectra
needed a 302-point Lebedev quadrature~\cite{Bouldi-22}.
For electron scattering spectra the situation is more complex and 
an average over orientations required
a 590-point Lebedev quadrature~\cite{Yang-21-2}.
Moreover, a Lebedev quadrature averages a function $f(\theta,\phi)$
of the two angles $\theta$ and $\phi$
whereas we need to average spectra $f(\alpha,\beta,\gamma)$ depending on the 
three Euler angles that describe the orientation of the sample. 

In this section, we write the RIXS spectrum of a powder as
a weighted sum of RIXS spectra of a single crystal 
(or oriented molecule) with different
incident and scattered polarizations. 
Since RIXS calculations
can be computationally expensive, we look for the minimum 
number of such spectra wich turns out to be 21, and even of 15 spectra if the 
polarization state of the scattered beam is not measured. 
In the next section,
we show that the symmetry of the sample further decreases
this number.

\subsection{Spherical tensor analysis}
We know~\cite{Juhin-14} that the polarization dependence of 
the RIXS cross-section of an isotropic (e.g. powder) sample could
entirely be described, in the electric dipole approximation, by three
parameters denoted $S_0$, $S_1$ and $S_2$, defined in Eq.~\eqref{defSgg}
and related
to $S^{gg0}_{L_0}$ of Ref.~\onlinecite{Juhin-14} by $S_g=\sqrt{2g+1}\,S^{gg0}$.

The general formula for arbitrary polarization of the incident
beam and polarization detection of the scattered beam is
\begin{eqnarray*}
	\sigma(\epsilon_s,\epsilon)  &=&
-\frac{S_1}{6}+\frac{S_2}{10}+
\Big(\frac{S_1}{6}+\frac{S_2}{10}\Big)|\epsilon\cdot\epsilon_s|^2
\\&&
+
\Big(\frac{S_0}{3}-\frac{S_2}{15}\Big)|\epsilon\cdot\epsilon^*_s|^2.
\end{eqnarray*}
We previously stressed~\cite{Juhin-14} that the 
polarization $\epsilon_s$
of the scattered beam is characterized by the measurement device
and several experimental setups were devised to explicitly measure
the polarization of the scattered
beam~\cite{Ishii-13,GaoPhD,Braicovich-14,Kim-18,Manzanillas-24}.
From the theoretical point of view~\cite{Ishii-11,GaoPhD}
the crystal analyzers with (or without)
explicit polarization detection can be considered
as a polarization filter and when
we say that the polarization is not measured,
we mean that this filter gives the same weight
to all polarization directions.

So, if the polarization of the scattered beam is not measured, 
the cross section becomes
\begin{eqnarray}
	\sigma(\widehat\bfk_s,\epsilon)  &=&
	C_0
	+C_1
	\big(|\epsilon\cdot\widehat\bfk_s|^2-1/3\big),
    \label{angdepks}
\end{eqnarray}
where $\widehat\bfk_s$ is the unit vector along the scattered beam and
\begin{eqnarray}
	C_0  =
	\frac{S_0}{9}-\frac{S_1}{9}+\frac{S_2}{9},\,\,\,
C_1=-
	\frac{S_0}{6}-\frac{S_1}{12}-\frac{S_2}{60}
	.
    \label{CzeroCun}
\end{eqnarray}
The term $-1/3$ in Eq.~\eqref{angdepks} is used to isolate $C_0$,
which is the RIXS cross-section when all scattered beams are 
collected~\cite{Juhin-14}. In other words, $C_0$ is the RIXS cross-section that would be measured with a detector (emission spectrometer) that would cover all directions in space.

The coefficients $S_b$ are defined by
\begin{eqnarray}
	S_b &=& \sum_{imjn} X_{im,jn}(b) a_{im,jn},
	\label{defSgg}
\end{eqnarray}
with
\begin{eqnarray}
	X_{im,jn}(b) &=&\sum_{\beta=-b}^b(-1)^{b-\beta}
\sum_{\mu\nu\mu'\nu'}
(1\mu'1\nu|b\beta)\nonumber\\&&\times(1\mu1\nu'|b-\beta)
A_{i\mu'} A_{m\nu} A_{n\mu}A_{j\nu'} ,
\label{defXimjn}
\end{eqnarray}
where the matrix elements of $A$ (defined in Eq.~\eqref{defmatriceA})
appear because the original definition of $S^{bb0}_{L_0}$ in~\cite{Juhin-14}
was in terms of spherical components $\bfr^{(1)}_\mu$ and we work
now with Cartesian components $r_n$. 
$X(b)$ is also related to the tensor of Section~\ref{Ydefsect}
by $X(b)=\sqrt{2b+1}Y^{bb}(0,0)$.
We changed the normalization to avoid the cluttering of
square roots in formulas and tables, but 
 $X(b)$ are no longer orthonormal for the Hilbert-Schmidt scalar product:
\begin{eqnarray}
	\sum_{imjn} X_{im,jn}(b)X^*_{im,jn}(b')  &=&
	  (2b+1)\delta_{bb'}.
	  \label{normXb}
\end{eqnarray}

Because of the definition of $X(b)$ as a coupling of spherical tensors, the matrix elements
$X_{im,jn}(b)$ enjoy the same relations as $a_{im,jn}$ for a spherical sample (see Section~\ref{sphersect}).
In addition to the permutation symmetries
discussed in Section~\ref{Ydefsect}, we note
that each $X(b)$ is real-valued and the three tensors $X(b)$ commute in three ways
\begin{eqnarray*}
	\sum_{kl} X_{im,kl}(b)X_{kl,jn}(b') &=& \sum_{kl} X_{im,kl}(b')X_{kl,jn}(b),\\
	\sum_{kl} X_{km,ln}(b)X_{km',ln'}(b') &=& \sum_{kl} X_{km,ln}(b')X_{km',ln'}(b),\\
	\sum_{kl} X_{ik,jl}(b)X_{i'k,j'l}(b') &=& \sum_{kl} X_{ik,jl}(b')X_{i'k,j'l}(b),
\end{eqnarray*}
and they are related to the spherical average of product of rotation matrices 
in Eq.~\eqref{invariancematsphe}
by
\begin{eqnarray*}
	N_{imjn,i'm'j'n'} &=& 
	\sum_{b=0}^2 \frac{1}{2b+1} X_{im,nj}(b)X_{i'm',n'j'}(b),
\end{eqnarray*}
which means that $X(b)$ are eigenstates of $N$ for the eigenvalue 1,
taking into account the normalization of Eq.~\eqref{normXb}.

We are now in a position to describe the problem at hand. 
We want to compute $S_b$ for $b=0,1,2$ and 
Eq.~\eqref{defSgg} suggests that, in the absence of symmetry, we need
to determine the 81 fundamental spectra of $a_{im,jn}$.
Moreover, some computer programs do not directly calculate
$a_{im,jn}$, but cross-sections $\sigma(\epsilon_s,\epsilon)$
related to $a_{im,jn}$ by Eq.~\eqref{defsigmaa}.
Thus, we wish to find a certain number $L$ of 
pairs of polarization vectors $(\epsilon_{sl},\epsilon_l)$ such that
\begin{eqnarray*}
		S_b &=& \sum_{l=1}^L
		\rho_l(b) \sigma(\epsilon_{sl},\epsilon_l),
\end{eqnarray*}
where $\rho_l(b)$ is the weight of $\sigma(\epsilon_{sl},\epsilon_l)$
for $S_b$ and we would like to minimize the number
$L$ of calculations.

As proved in Eq.~\eqref{polRIXS}, a generalization of the polarization identity
requires 16 calculations of $\sigma(\epsilon_s,\epsilon)$
to calculate each $a_{im,jn}$. Therefore, a brute force approach
would require $L=16\times 81=1296$ calculations of cross-sections to compute
the three parameters of the RIXS of a powder. 
We shall reduce this number to $L=21$.

\subsection{Generalized polarization identity}
\label{polidsect}
The polarization identity is a relation between 
\begin{eqnarray*}
	f(\bfu,\bfv) &=& \sum_{mn} u_m^* a_{mn} v_n,
\end{eqnarray*}
where $a_{mn}$ is a (finite dimensional) complex matrix,
and $\phi(\bfu)=f(\bfu,\bfu)$. Namely
\begin{eqnarray*}
	f(\bfu,\bfv) &=& \frac{1}{4} \sum_{k=0}^3 (-i)^k\phi(\bfu+i^k\bfv).
\end{eqnarray*}

Our case is more complicated because we must define
\begin{eqnarray}
	f(\bfu,\bfu';\bfv,\bfv') &=& \sum_{imjn}
	u_i^* v_m^*  a_{im,jn} u'_j v'_n,
	\label{fuupvvp}
\end{eqnarray}
and $\phi(\bfu,\bfv)=f(\bfu,\bfu;\bfv,\bfv)$, so that
$\sigma(\epsilon_s,\epsilon)=\phi(\epsilon_s,\epsilon^*)$.
A lengthy but straightforward calculation proves the
generalized polarization identity that we need:
\begin{eqnarray}
	f(\bfu,\bfu';\bfv,\bfv') &=&  \frac{1}{16}
	\sum_{m=0}^3 \sum_{n=0}^3
	(-i)^{m+n}
\nonumber\\&&\times
	\phi(\bfu+i^m\bfu';\bfv+i^n\bfv').
	\label{polRIXS}
\end{eqnarray}
An immediate consequence of this identity is that
the invariance of the cross-section (and therefore of $\phi$) implies the invariance of $f$.
Indeed, Eq.~\eqref{actionung} implies
\begin{eqnarray*}
	\phi(g\bfu+i^m g\bfu';g\bfv'+i^n g\bfv')
	\\&&\hspace{-18mm}= \phi\big(g(\bfu+i^m\bfu');g(\bfv'+i^n \bfv')\big)
	\\&&\hspace{-18mm}=
	\phi(\bfu+i^m \bfu';\bfv'+i^n \bfv'),
\end{eqnarray*}
and then, by the polarization identity,
\begin{eqnarray}
	f(g\bfu,g\bfu';g\bfv,g\bfv')
	&=& f(\bfu,\bfu';\bfv,\bfv'),
	\label{invariancef}
\end{eqnarray}
for every $g$ in the invariance group $G$ of the sample.
This identity extends the invariance of the function $\sigma(\epsilon_s,\epsilon)$ 
of two vectors $\epsilon_s$ and $\epsilon$ to the invariance of a function
$f$ of four vectors. 

Similarly, Eq.~\eqref{Rtrianglesigma} implies that, for any isometry $g$
\begin{eqnarray}
(g^{-1}\triangleright f)(\bfu,\bfu';\bfv,\bfv') &=&f(g\bfu,g\bfu';g\bfv,g\bfv').
	\label{gtrianglef}
\end{eqnarray}

\subsection{Symmetrization}
We are now ready to prove that the left hand
side of Eq.~\eqref{invariancea}
has all the symmetries of the group $G$ even if $ a_{i'm',j'n'}$
has no symmetry. 
More generally, we prove that, if $f$ has no invariance whatsoever,
then the average $\overline{f}$ defined by
\begin{eqnarray*}
\overline{f}(\bfu,\bfu';\bfv,\bfv')	&=& \frac{1}{|G|}\sum_{g\in G} f(g\bfu,g\bfu';g\bfv,g\bfv').
\end{eqnarray*}
is invariant under any operation $g$ of the invariance group $G$ of the sample.
This statement will be clear to the group theory experts, because the
right hand side is just the projection onto the fully symmetric representation of $G$,
but let us show it. Let $h\in G$ be an element of the symmetry group. 
By linearity of the action
\begin{eqnarray*}
	(h^{-1}\triangleright\overline{f})(\bfu,\bfu';\bfv,\bfv')= \sum_{g\in G} \frac{(h^{-1}\triangleright f)(g\bfu,g\bfu';g\bfv,g\bfv')}{|G|}.
\end{eqnarray*}
Now we can use Eq.~\eqref{gtrianglef}, which is valid even if $h$ is not
a symmetry of $f$:
\begin{eqnarray*}
	(h^{-1}\triangleright\overline{f})(\bfu,\bfu';\bfv,\bfv')= \sum_{g\in G} \frac{f(hg\bfu,hg\bfu';hg\bfv,hg\bfv')}{|G|}.
\end{eqnarray*}
Any element $g$ can be written in a unique way as $g=h^{-1}g'$ and $g'$ runs over
$G$ when $g$ runs over $G$. Thus, 
\begin{eqnarray*}
	(h^{-1}\triangleright\overline{f})(\bfu,\bfu';\bfv,\bfv')&= &\sum_{g'\in G} \frac{f(g'\bfu,g'\bfu';g'\bfv,g'\bfv')}{|G|}
	\\&=& \overline{f}(\bfu,\bfu';\bfv,\bfv').
\end{eqnarray*}
Since this is true for any $h\in G$,
 $\overline{f}$ is invariant under $G$ even if $f$ is not. 
 Besides, if $f$ is already invariant under $G$, then
 $\overline{f}=f$.

\subsection{Singular value decomposition}
To decrease the number of calculations, we 
write a sort of singular value decomposition of $X(b)$:
\begin{eqnarray}
	X_{im,jn}(b) &=& \sum_{k=1}^9 \lambda^k(b) U^k_{ij} (V^k_{mn})^*,
	\label{XUoV}
\end{eqnarray}
where it will turn out that the $3\times 3$ matrices $U^k$ and $V^k$
do not depend on $b$. If 
we can diagonalize $U^k$ and $V^k$ as
\begin{eqnarray*}
	U^k_{ij} &=& \sum_{r=1}^3 \mu^{kr} u^{kr}_i (u^{kr}_j)^*,\\
	V^k_{mn} &=& \sum_{s=1}^3 \nu^{ks} v^{ks}_m (v^{ks}_n)^*,
\end{eqnarray*}
we obtain by definition~\eqref{defSgg} of $S_b$:
\begin{eqnarray}
S_b	&=& \sum_{k,r,s}
	\lambda^k(b) \mu^{kr} \nu^{ks} 
	\phi((\bfu^{kr})^* ,\bfv^{ks})).
	\label{Sbb}
\end{eqnarray}

Equation~\eqref{Sbb} is already an improvement because the 
sum over $k,r,s$ involves 81 cross-sections 
$\phi((\bfu^{kr})^* ,\bfv^{ks})$
instead of 1296, but further progress is possible. 

To implement this singular value decomposition, we define three $9\times 9$ matrices
$M(b)$ by $M(b)_{3(i-1)+j,3(m-1)+n}=X_{im,jn}(b)$:
\begin{eqnarray*}
	M(0) &=& \frac{1}{3} \id_9,
\end{eqnarray*}
\begin{eqnarray*}
	M(1) &=& \frac{1}{2}
	\left(
	\begin{array}{ccccccccc}
		0 & 0 & 0 & 0 & -1 & 0 & 0 & 0 & -1 \\
		0 & 0 & 0 & 1 & 0 & 0 & 0 & 0 & 0 \\
		0 & 0 & 0 & 0 & 0 & 0 & 1 & 0 & 0 \\
		0 & 1 & 0 & 0 & 0 & 0 & 0 & 0 & 0 \\
		-1 & 0 & 0 & 0 & 0 & 0 & 0 & 0 & -1 \\
		0 & 0 & 0 & 0 & 0 & 0 & 0 & 1 & 0 \\
		0 & 0 & 1 & 0 & 0 & 0 & 0 & 0 & 0 \\
		0 & 0 & 0 & 0 & 0 & 1 & 0 & 0 & 0 \\
		-1 & 0 & 0 & 0 & -1 & 0 & 0 & 0 & 0 \\
	\end{array}
	\right),
\end{eqnarray*}
and
\begin{eqnarray*}
	M(2) &=& \frac{1}{6}
	\left(
	\begin{array}{ccccccccc}
		4 & 0 & 0 & 0 & 3 & 0 & 0 & 0 & 3 \\
		0 &-2 & 0 & 3 & 0 & 0 & 0 & 0 & 0 \\
		0 & 0 & -2 & 0 & 0 & 0 & 3 & 0 & 0 \\
		0 & 3 & 0 & -2 & 0 & 0 & 0 & 0 & 0 \\
		3 & 0 & 0 & 0 & 4 & 0 & 0 & 0 & 3 \\
		0 & 0 & 0 & 0 & 0 & -2 & 0 & 3 & 0 \\
		0 & 0 & 3 & 0 & 0 & 0 & -2 & 0 & 0 \\
		0 & 0 & 0 & 0 & 0 & 3 & 0 & -2 & 0 \\
		3 & 0 & 0 & 0 & 3 & 0 & 0 & 0 & 4 \\
	\end{array}
	\right).
\end{eqnarray*}
These matrices commute and are real symmetric, so 
the singular value decomposition turns out
to be a diagonalization of $M(b)$ (this would not be 
the case if quadrupole transitions were taken
into account) and a common set of real ``eigenvectors'' $U^k$ and
$V^k=U^k$ can be chosen for the three $X(b)$. 
Since all $U^k$ are real we can write
\begin{eqnarray*}
	X_{im,jn}(b) &=&
	\sum_{krs}
	\lambda^k(b) \mu^{kr} \mu^{ks} u^{kr}_i u^{ks}_m 
	(u^{kr}_j)^*(u^{ks}_n)^*.
\end{eqnarray*}
We further observe in Table~\ref{vecteurbrut} that each $U^k$ is either symmetric
or antisymmetric, so the eigenvalues $\mu^{kr}$ for
a given $k$ are either all real or all purely imaginary and
\begin{eqnarray}
	X_{im,jn}(b)=
	\sum_{krs}
	\lambda^k(b) \mu^{kr} \mu^{ks} (u^{kr}_i)^* (u^{ks}_m)^* 
	u^{kr}_j u^{ks}_n.
	\label{Xsum33}
\end{eqnarray}
Thus,
\begin{eqnarray*}
	S_b	&=& \sum_{krs}
	\lambda^k(b) \mu^{kr} \mu^{ks} 
	\phi(\bfu^{kr} ,\bfu^{ks}).
\end{eqnarray*}

\begin{table}
	\caption{Eigenvalues $\lambda^k(b)$ of $M(b)$ for $b=0,1,2$ corresponding to
		their common eigenvectors $U^k$.
		The last line indicates whether
		$U^k$ is symmetric ($+$)
		or antisymmetric $(-)$.
	}
\label{vecteurbrut}
\begin{center}
	\bgroup
	\def\arraystretch{1.3}
	\begin{tabular}{|c|c|c|c|c|c|c|c|c|c|} \hline
		$b$ & $U^1$ & $U^2$ & $U^3$ & $U^4$ & 
		$U^5$ & $U^6$ & $U^7$ & $U^8$ & $U^9$\\
		\hline 
		$0$ & $\frac{1}{3}$ & $\frac{1}{3}$ & $\frac{1}{3}$ & $\frac{1}{3}$ & 
		$\frac{1}{3}$ & $\frac{1}{3}$ & $\frac{1}{3}$ & $\frac{1}{3}$ & $\frac{1}{3}$\\
		\hline 
		$1$ & $-1$ & $-\frac{1}{2}$ & $-\frac{1}{2}$ & $-\frac{1}{2}$ & 
		$\frac{1}{2}$ & $\frac{1}{2}$ & $\frac{1}{2}$ & $\frac{1}{2}$ & $\frac{1}{2}$\\
		\hline 
		$2$ & $\frac{5}{3}$ & $-\frac{5}{6}$ & $-\frac{5}{6}$ & $-\frac{5}{6}$ & 
		$\frac{1}{6}$ & $\frac{1}{6}$ & $\frac{1}{6}$ & $\frac{1}{6}$ & $\frac{1}{6}$\\
		\hline 
		& $+$ & $-$ & $-$ & $-$ & 
		$+$ & $+$ & $+$ & $+$ & $+$\\
		\hline 
	\end{tabular}
	\egroup
	\vskip 3mm
\end{center}
\end{table}
The eigenvalues $\lambda^k(b)$ are given in Table~\ref{vecteurbrut}
and the nine matrices $U^k$ are
\begin{eqnarray*}
U^1 &= & \frac{1}{\sqrt{3}}\left(
	\begin{array}{ccc}
		1 & 0 & 0 \\
		0 & 1 & 0 \\
		0 & 0 & 1 \\
	\end{array}
	\right),
	\quad U^2 = \frac{1}{\sqrt{2}}
	\left(
	\begin{array}{ccc}
		0 & 0 & 0 \\
		0 & 0 & -1 \\
		0 & 1 & 0 \\
	\end{array}
	\right),
\end{eqnarray*}
\begin{eqnarray*}
U^3 &=& \frac{1}{\sqrt{2}}
	\left(
	\begin{array}{ccc}
		0 & 0 & -1 \\
		0 & 0 & 0 \\
		1 & 0 & 0 \\
	\end{array}
	\right),
	\quad U^4 = \frac{1}{\sqrt{2}}
	\left(
	\begin{array}{ccc}
		0 & -1 & 0 \\
		1 & 0 & 0 \\
		0 & 0 & 0 \\
	\end{array}
	\right),
\end{eqnarray*}
\begin{eqnarray*}
	U^5 &=& \frac{1}{\sqrt{2}}
	\left(
	\begin{array}{ccc}
		-1 & 0 & 0 \\
		0 & 0 & 0 \\
		0 & 0 & 1 \\
	\end{array}
	\right),
	\quad U^6 =  \frac{1}{\sqrt{2}}
	\left(
	\begin{array}{ccc}
		0 & 0 & 0 \\
		0 & 0 & 1 \\
		0 & 1 & 0 \\
	\end{array}
	\right),
\end{eqnarray*}
\begin{eqnarray*}
	U^7 &=& \frac{1}{\sqrt{2}}
	\left(
	\begin{array}{ccc}
		0 & 0 & 1 \\
		0 & 0 & 0 \\
		1 & 0 & 0 \\
	\end{array}
	\right),
	\quad U^8 =\frac{1}{\sqrt{6}}
	\left(
	\begin{array}{ccc}
		-1 & 0 & 0 \\
		0 & 2 & 0 \\
		0 & 0 & -1 \\
	\end{array}
	\right),
\end{eqnarray*}
\begin{eqnarray*}
	U^9 &=& \frac{1}{\sqrt{2}}
	\left(
	\begin{array}{ccc}
		0 & 1 & 0 \\
		1 & 0 & 0 \\
		0 & 0 & 0 \\
	\end{array}
	\right).
\end{eqnarray*}

Matrices $U^k$ for $k=1, 5$ and $8$ are already diagonal, so 
we can take for eigenvectors $\bfu^{kr}$ (for $k=1, 5$ and $8$) the same
real unit vectors 
$\bfe_x$, $\bfe_y$ and $\bfe_z$ along the $x$, $y$ and $z$ axes,
respectively.
This means that we can calculate $k=1, 5$ and $8$ with the same 9 values of
$\phi(\bfe_p ,\bfe_q)$, where $p$ and $q$ take the values $x,y$ and $z$, even if only four of them are useful for $k=5$
because of its zero eigenvalue.
 
For $k=2,3$ and $4$, each $U^k$ is a real antisymmetric matrix, its two non-zero eigenvalues
are $\mu^{kr}=\pm \imath/\sqrt2$, the corresponding eigenvectors are
of the form $\bfu^{kr}=\bfpi_{pq}^\pm = (\pm \imath\bfe_p+\bfe_q)/\sqrt2$
(i.e. $\bfpi_{yz}^\pm$ for $k=2$, $\bfpi_{xz}^\pm$ for $k=3$, $\bfpi_{xy}^\pm$ for $k=4$)
and we have to calculate 4 different $\phi(\bfu^{kr} ,\bfu^{ks})$ for each $k$.

For $k=6, 7$ and $9$, each $U^k$ is a real symmetric matrix,
its two non-zero eigenvalues
are $\mu^{kr}=\pm 1/\sqrt2$, the corresponding eigenvectors are
of the form $\bfu^{kr}=\bfe_{pq}^\pm = (\pm \bfe_p+\bfe_q)/\sqrt2$
(i.e. $\bfe_{yz}^\pm$ for $k=6$, $\bfe_{xz}^\pm$ for $k=7$, $\bfe_{xy}^\pm$ for $k=9$)
and we have again to calculate 4 different $\phi(\bfu^{kr} ,\bfu^{ks})$
for each $k$.

Therefore, the total number of cross-sections to calculate is
reduced from 81 to
$9+3\times 4+3\times 4=33$.

\subsection{Further reduction}
To proceed with the reduction, we define 
the fourth-rank tensors $(\bfu,\bfv)$ with
components
\begin{eqnarray*}
	(\bfu,\bfv)_{imjn} &=& u_i^* v_m^* u_j v_n,
\end{eqnarray*}
and we write Eq.~\eqref{Xsum33} in a more general form:
\begin{eqnarray}
	X_{im,jn}(b) =
	\sum_{l}
	\rho_l(b) (\bfu_l,\bfv_l)_{imjn},	\label{Xsumn}
\end{eqnarray}
where $l$ enumerates the $krs$ indices, 
$(\bfu_l,\bfv_l)$ is a fourth-rank tensor
and $\rho_l(b) \in\bbR$ is its weight.
After reduction, Eq.~\eqref{Xsumn} will
still be valid whereas Eq.~\eqref{Xsum33} will not.

By examining the 33 fourth-rank tensors in Eq.~\eqref{Xsum33},
we observe the following linear relations
\begin{eqnarray*}
\sum_{r,s\in\{p,q\}} (\bfe_r,\bfe_s) &=&
    \sum_{\eta,\eta'\in\{+,-\}}(\bfpi^\eta_{pq},\bfpi^{\eta'}_{pq}),\\
\sum_{r,s\in\{p,q\}} (\bfe_r,\bfe_s) &=&  \sum_{\eta,\eta'\in\{+,-\}}(\bfe^\eta_{pq},\bfe^{\eta'}_{pq}),
\end{eqnarray*}
where $(p,q)=(y,z), (x,z)$ and $(x,y)$.
These 6 linear relations would enable us to remove
six fourth-rank tensors $(\bfu_l,\bfv_l)$.
However, we can further reduce them by
observing that  $(\bfu^+,\bfu^-)$ and $(\bfu^-,\bfu^+)$ always have the same
weight in the calculation of $X(b)$ by Eq.~\eqref{Xsum33}
for any $\bfu=\bfpi_{pq}$ and $\bfu=\bfe_{pq}$. 
Therefore, we can use
\begin{eqnarray*}
	\sum_{\eta=\pm}(\bfpi^\eta_{pq},\bfpi^{-\eta}_{pq})
	&=& \sum_{r,s\in\{p,q\}} (\bfe_r,\bfe_s)
	-\sum_{\eta=\pm}(\bfpi^\eta_{pq},\bfpi^\eta_{pq}),\\
	\sum_{\eta=\pm}(\bfe^\eta_{pq},\bfe^{-\eta}_{pq})
&=& \sum_{r,s\in\{p,q\}} (\bfe_r,\bfe_s)
-\sum_{\eta=\pm}(\bfe^\eta_{pq},\bfe^\eta_{pq}),
\end{eqnarray*}
to remove 12 tensors $(\bfu_l,\bfv_l)$ instead of 6 and we are finally left with
21 fourth-rank tensors.

Our final result is 
\begin{eqnarray}
	S_b	= \sum_{l=1}^{21}
	\rho_l(b) \phi\big(\bfu_l,\bfv_l\big)=
	\sum_{l=1}^{21}
	\rho_l(b) \sigma\big(\bfu_l,\bfv_l^*\big),
	\label{finalSgg0}
\end{eqnarray}
for the tensors and the weights given in Table~\ref{poidsbase}.

\begin{table}
	\caption{Value of the weight $\rho_l(b)$ of  
		$\sigma(\bfu_l, \bfv_l^*)$ for the minimal 21 fourth-rank tensors
		$(\bfu_l,\bfv_l)$
		 to calculate $S^{bb0}$ by Eq.~\eqref{finalSgg0}
         as well as $C_0$ and $C_1$ in Eq.~\eqref{angdepks}.
		The basis unit vectors $\bfe_x$, $\bfe_y$ and $\bfe_z$
		are along the three axes of $\bbR^3$,
		$\bfpi_{mn}^\pm = (\pm i\bfe_m+\bfe_n)/\sqrt2$
		and
		$\bfe_{mn}^\pm = (\pm \bfe_m+\bfe_n)/\sqrt2$.
	}
\label{poidsbase}
\begin{center}
	\bgroup
	\def\arraystretch{1.3}
	\begin{tabular}{|c|c|c|c|c|c|c|} \hline
		$l$ & $(\bfu_l,\bfv_l)$ & $\rho_l(0)$ & $\rho_l(1)$ & $\rho_l(2)$ & $C_0$ & $C_1$\\
		\hline
		1 & $(\bfe_x,\bfe_x)$ & $1/3$ & $-1$ & $-1/3$ & 1/9 & 1/30\\
		\hline
		2 & $(\bfe_x,\bfe_y)$ & $0$ & $-1$ & $0$ & 1/9 & 1/12 \\
		\hline
		3 & $(\bfe_x,\bfe_z)$ & $0$ & $-1$ & $0$ & 1/9 & 1/12  \\
		\hline
		4 & $(\bfe_y,\bfe_x)$ & $0$ & $-1$ & $0$ & 1/9 & 1/12  \\
		\hline
		5 & $(\bfe_y,\bfe_y)$ & $1/3$ & $-1$ & $-1/3$ & 1/9 & 1/30 \\
		\hline
		6 & $(\bfe_y,\bfe_z)$ & $0$ & $-1$ & $0$ & 1/9 & 1/12  \\
		\hline
		7 & $(\bfe_z,\bfe_x)$ & $0$ & $-1$ & $0$ & 1/9 & 1/12  \\
		\hline
		8 & $(\bfe_z,\bfe_y)$ & $0$ & $-1$ & $0$ & 1/9 & 1/12  \\
		\hline
		9 & $(\bfe_z,\bfe_z)$ & $1/3$ & $-1$ & $-1/3$ & 1/9 & 1/30 \\
		\hline
		10 & $(\bfpi_{yz}^+,\bfpi_{yz}^+)$ & $-1/3$ & $1/2$ & $5/6$ & 0 & 0 \\
		\hline
		11 & $(\bfpi_{yz}^-,\bfpi_{yz}^-)$ & $-1/3$ & $1/2$ & $5/6$ & 0 & 0 \\
		\hline
		12 & $(\bfpi_{xz}^+,\bfpi_{xz}^+)$ & $-1/3$ & $1/2$ & $5/6$ & 0 & 0 \\
		\hline
		13 & $(\bfpi_{xz}^-,\bfpi_{xz}^-)$ & $-1/3$ & $1/2$ & $5/6$ & 0 & 0 \\
		\hline
		14 & $(\bfpi_{xy}^+,\bfpi_{xy}^+)$ & $-1/3$ & $1/2$ & $5/6$ & 0 & 0 \\
		\hline
		15 & $(\bfpi_{xy}^-,\bfpi_{xy}^-)$ & $-1/3$ & $1/2$ & $5/6$ & 0 & 0 \\
		\hline
		16 & $(\bfe_{yz}^+,\bfe_{yz}^+)$ & $1/3$ & $1/2$ & $1/6$ & 0 & -1/10 \\
		\hline
		17 & $(\bfe_{yz}^-,\bfe_{yz}^-)$ & $1/3$ & $1/2$ & $1/6$ & 0 & -1/10 \\
		\hline
		18 & $(\bfe_{xz}^+,\bfe_{xz}^+)$ & $1/3$ & $1/2$ & $1/6$ & 0 & -1/10 \\
		\hline
		19 & $(\bfe_{xz}^-,\bfe_{xz}^-)$ & $1/3$ & $1/2$ & $1/6$ & 0 & -1/10 \\
		\hline
		20 & $(\bfe_{xy}^+,\bfe_{xy}^+)$ & $1/3$ & $1/2$ & $1/6$ & 0 & -1/10 \\
		\hline
		21 & $(\bfe_{xy}^-,\bfe_{xy}^-)$ & $1/3$ & $1/2$ & $1/6$ & 0 & -1/10 \\
		\hline
	\end{tabular}
	\egroup
	\vskip 3mm
\end{center}
\end{table}

We can also write $C_0$ and $C_1$ 
in the angular dependence of Eq.~\eqref{angdepks}
directly in terms of 15 cross-sections:
\begin{eqnarray}
	C_0 &=& \frac{1}{9} \sum_{pq} \sigma(\bfe_p,\bfe_q),\label{Czerogeneral}\\
    C_1 &=& \frac{1}{12} \sum_{pq} \sigma(\bfe_p,\bfe_q)
    - \frac{1}{20} \sum_{p} \sigma(\bfe_p,\bfe_p)
    \nonumber\\&&
    - \frac{1}{10} \sum_{p<q} \big(\sigma(\bfe^+_{pq},\bfe^+_{pq})
    + \sigma(\bfe^-_{pq},\bfe^-_{pq})\big),\label{Cungeneral}
\end{eqnarray}
where $p$ and $q$ run over $\{x,y,z\}$ and
$p<q$ over $\{xy,yz,xz\}$.

\section{RIXS for a powder with symmetry}
\label{sympowsect}

We can now combine the discussion of the influence of the symmetry of the
sample and the number of $\phi(\epsilon_s,\epsilon)$ required
to calculate the spectrum of a powdered material when the
material itself has a symmetry.

The basic idea is quite simple. If $G$ is the symmetry group
of the material, we compute $\phi(\bfu_l,\bfv_l)$, where
\begin{eqnarray*}
	\phi(\bfu_l,\bfv_l) &=& 
	\sum_{imjn} u_{li}^*v_{lm}^* 
	 a_{im,jn}u_{lj} v_{ln},
\end{eqnarray*}
for the 21 pairs of polarizations $(\bfu_l,\bfv_l)$ of Table~\ref{poidsbase}
in terms of the $n_G$ fundamental spectra
of $a_{im,jn}$.
This gives us a linear relation between $\phi(\bfu_l,\bfv_l)$
and the fundamental spectra described by
a $n_G\times 21$ matrix whose null space represents
the relations between different $\phi(\bfu_l,\bfv_l)$.
For the example of a cubic material, the expression
of $\phi(\bfu_l,\bfv_l)$ in terms of the four
fundamental spectra $a_{11,11}$, $a_{11,22}$,
$a_{12,12}$ and $a_{12,21}$ is
\begin{eqnarray*}
	\phi(\bfe_p,\bfe_q) &=& a_{12,12}+ \delta_{pq}(a_{11,11}-a_{12,12}),\\ 
	\phi(\bfpi^\pm_{pq},\bfpi^\pm_{pq}) &=& \frac12(a_{11,11}-a_{11,22}+a_{12,12}+a_{12,21}),\\ 
	\phi(\bfe^\pm_{pq},\bfe^\pm_{pq}) &=& \frac12(a_{11,11}+a_{11,22}+a_{12,12}+a_{12,21}),
\end{eqnarray*}
for $p$ and $q$ equal to $x,y$ and $z$. The relations between $\phi(\bfu_l,\bfv_l)$
is then $\phi(\bfe_p,\bfe_p)=\phi(\bfe_x,\bfe_x)$ for all $p$, and
for all $q\not=p$:
$\phi(\bfe_p,\bfe_q)=\phi(\bfe_x,\bfe_y)$,
$\phi(\bfpi^\pm_{pq},\bfpi^\pm_{pq})=\phi(\bfpi^\pm_{xy},\bfpi^\pm_{xy})$
 and
$\phi(\bfe^\pm_{pq},\bfe^\pm_{pq})=\phi(\bfe^\pm_{xy},\bfe^\pm_{xy})$.
For a powder of a cubic sample only four calculations are required to compute
the full angular dependence of RIXS. This number reduces to three when
the polarization state of the scattered beam is not measured,
because $\phi(\bfpi^\pm_{pq},\bfpi^\pm_{pq})$ is not needed.

We list now the fundamental spectra required to calculate
the RIXS spectum of a powder of a sample with various symmetry groups.
For all crystallographic point groups,
the minimal number of spectra $n_S$ and $n_C$ needed 
to compute $S_b$ and ($C_0, C_1)$, respectively, is
given in Table~\ref{crystptgroup}.

\subsection{For $O$, $T_d$ and $O_h$ symmetry groups}
In the previous paragraph, we showed that 
$\phi(\bfu_l,\bfu_v)$ can take only four different values.
Thus, Eq.~\eqref{finalSgg0} becomes:

\begin{eqnarray*}
	S_0 &=& \sigma(\bfe_x,\bfe_x) 
	+ 2 \sigma(\bfe^+_{xy},\bfe^+_{xy}) -2 \sigma(\bfpi^+_{xy},\bfpi^-_{xy}),\\
	S_1 &=& -3\sigma(\bfe_x,\bfe_x)-6\sigma(\bfe_x,\bfe_y)
	\\&& +3\sigma(\bfe^+_{xy},\bfe^+_{xy}) 
      +3\sigma(\bfpi^+_{xy},\bfpi^-_{xy}),\\
	S_2 &=& -\sigma(\bfe_x,\bfe_x) 
	+ \sigma(\bfe^+_{xy},\bfe^+_{xy}) +5\sigma(\bfpi^+_{xy},\bfpi^-_{xy}).
\end{eqnarray*}

The most general polarization dependence of a powder of cubic sample can
be obtained from the calculation of four spectra:
\begin{eqnarray*}
	\sigma(\epsilon_s,\epsilon) &=& \big(\sigma(\bfe^+_{xy},\bfe^+_{xy}) -\sigma(\bfpi^+_{xy},\bfpi^-_{xy})\big)
	|\epsilon\cdot\epsilon_s^*|^2
	\\&+&
	\big(\sigma(\bfe^+_{xy},\bfe^+_{xy}) +\sigma(\bfpi^+_{xy},\bfpi^-_{xy})
	\\&&\hspace{6mm}
	-\sigma(\bfe_x,\bfe_x) -\sigma(\bfe_x,\bfe_y) \big)|\epsilon\cdot\epsilon_s|^2
	\nonumber\\&+&
	\sigma(\bfe_x,\bfe_y) ||\epsilon||^2||\epsilon_s||^2.
\end{eqnarray*}

For polarization vectors normalized to unity and when the polarization
of the scattered beam is not measured, three calculations are enough:
\begin{eqnarray*}
	C_0 &=& \frac{\sigma(\bfe_x,\bfe_x)+2\sigma(\bfe_x,\bfe_y)}{3},\\
    C_1 &=& \frac{\sigma(\bfe_x,\bfe_x)+5\sigma(\bfe_x,\bfe_y)-6\sigma(\bfe^+_{xy},\bfe^+_{xy})}{10}.
\end{eqnarray*}

\subsection{For $D_3$, $D_{3d}$,  $C_{3v}$, $D_6$, $C_{6v}$, $D_{3h}$ and $D_{6h}$}
For the groups $D_3$, $D_{3d}$ or  $C_{3v}$, the relations between
$\phi(\bfu_l,\bfv_l)$ are not the same as those for
$D_6$, $C_{6v}$, $D_{3h}$ or $D_{6h}$, but the differences compensate
each other and the same final formulas for $S_b$ are obtained.
Indeed, the relations between $\phi(\bfu_l,\bfv_l)$ for $D_3$ are:
\begin{eqnarray*}
\phi(\bfe^+_{xz},\bfe^+_{xz}) &=& \frac12\big(\phi(\bfe^-_{yz},\bfe^-_{yz})+\phi(\bfe^+_{yz},\bfe^+_{yz})\big),\\
\phi(\bfe^-_{xz},\bfe^-_{xz}) &=& \phi(\bfe^+_{xz},\bfe^+_{xz}),\\
\phi(\bfe^-_{xy},\bfe^-_{xy}) &=&
\phi(\bfe^+_{xy},\bfe^+_{xy}) =
\phi(\bfe_y,\bfe_y) = \phi(\bfe_x,\bfe_x),\\
\phi(\bfpi^+_{xz},\bfpi^+_{xz}) &=& \frac12\big(\phi(\bfpi^-_{yz},\bfpi^-_{yz}) +
\phi(\bfpi^+_{yz},\bfpi^+_{yz})\big),\\
\phi(\bfpi^-_{xz},\bfpi^-_{xz}) &=& \phi(\bfpi^+_{xz},\bfpi^+_{xz}),\\
\phi(\bfpi^-_{xy},\bfpi^-_{xy}) &=& \phi(\bfpi^+_{xy},\bfpi^+_{xy}),\\
\phi(\bfe_z,\bfe_y) &=& \phi(\bfe_z,\bfe_x),\\
\phi(\bfe_y,\bfe_z) &=& \phi(\bfe_x,\bfe_z),\\
\phi(\bfe_y,\bfe_x) &=& \phi(\bfe_x,\bfe_y),
\end{eqnarray*}
while for $D_6$ they are
\begin{eqnarray*}
\phi(\bfe^-_{yz},\bfe^-_{yz}) &=& 
\phi(\bfe^+_{yz},\bfe^+_{yz}) = 
\phi(\bfe^-_{xz},\bfe^-_{xz}) = \phi(\bfe^+_{xz},\bfe^+_{xz}),\\
\phi(\bfpi^-_{yz},\bfpi^-_{yz}) &=& 
\phi(\bfpi^+_{yz},\bfpi^+_{yz}) = 
\phi(\bfpi^-_{xz},\bfpi^-_{xz}) = \phi(\bfpi^+_{xz},\bfpi^+_{xz}),\\
\phi(\bfe^-_{xy},\bfe^-_{xy}) &=&
\phi(\bfe^+_{xy},\bfe^+_{xy}) =
\phi(\bfe_y,\bfe_y) = \phi(\bfe_x,\bfe_x),\\
\phi(\bfpi^-_{xy},\bfpi^-_{xy}) &=& \phi(\bfpi^+_{xy},\bfpi^+_{xy}),\\
\phi(\bfe_z,\bfe_y) &=& \phi(\bfe_z,\bfe_x),\\
\phi(\bfe_y,\bfe_z) &=& \phi(\bfe_x,\bfe_z),\\
\phi(\bfe_y,\bfe_x) &=& \phi(\bfe_x,\bfe_y).
\end{eqnarray*}

When the polarization state of the scattered beam is not measured
we need six calculations:
\begin{eqnarray*}
	C_0 &=& \frac{2}{9}\big(\sigma(\bfe_x,\bfe_x)+\sigma(\bfe_x,\bfe_y)+\sigma(\bfe_x,\bfe_z)+\sigma(\bfe_z,\bfe_x)\big)
   \\&&+\frac{\sigma(\bfe_z,\bfe_z)}{9}
    ,\\
    C_1 &=& -\frac{2}{15}\sigma(\bfe_x,\bfe_x)+\frac{\sigma(\bfe_x,\bfe_y)+\sigma(\bfe_x,\bfe_z)+\sigma(\bfe_z,\bfe_x)}{6}
   \\&&+\frac{\sigma(\bfe_z,\bfe_z)}{30}-\frac{2}{5}\sigma(\bfe^+_{xz},\bfe^+_{xz}).
\end{eqnarray*}
For the full spectra we need eight calculations.
\begin{eqnarray*}
S_0 &=& \frac{4}{3}\sigma(\bfe_x,\bfe_x)+\frac{1}{3}\sigma(\bfe_z,\bfe_z)
-\frac{2}{3}\sigma(\bfpi^+_{xy},\bfpi^-_{xy})
\\&&
-\frac{4}{3}\sigma(\bfpi^+_{xz},\bfpi^-_{xz})+\frac{4}{3}\sigma(\bfe^+_{xz},\bfe^+_{xz}),\\
S_1 &=& -\sigma(\bfe_x,\bfe_x)-2\sigma(\bfe_x,\bfe_y)-2\sigma(\bfe_x,\bfe_z)
\\&& -2 \sigma(\bfe_z,\bfe_x)
-\sigma(\bfe_z,\bfe_z)+\sigma(\bfpi^+_{xy},\bfpi^-_{xy})
\\&& +2\sigma(\bfpi^+_{xz},\bfpi^-_{xz})
+2\sigma(\bfe^+_{xz},\bfe^+_{xz}),\\
S_2 &=& -\frac{1}{3}\sigma(\bfe_x,\bfe_x)-\frac{1}{3}\sigma(\bfe_z,\bfe_z)+\frac{5}{3}\sigma(\bfpi^+_{xy},\bfpi^-_{xy})
\\&&
+\frac{10}{3}\sigma(\bfpi^+_{xz},\bfpi^-_{xz})+\frac{2}{3}\sigma(\bfe^+_{xz},\bfe^+_{xz}).
\end{eqnarray*}

\subsection{For $D_4$, $C_{4v}$, $D_{2d}$ and  $D_{4h}$}
For the groups $D_4$ or $C_{4v}$ or $D_{2d}$ or  $D_{4h}$
the relations between $\phi(\bfu_l,\bfv_l)$ are
\begin{eqnarray*}
\phi(\bfe^-_{yz},\bfe^-_{yz}) &=& 
\phi(\bfe^+_{yz},\bfe^+_{yz}) = 
\phi(\bfe^-_{xz},\bfe^-_{xz}) = \phi(\bfe^+_{xz},\bfe^+_{xz}),\\
\phi(\bfpi^-_{yz},\bfpi^-_{yz}) &=& 
\phi(\bfpi^+_{yz},\bfpi^+_{yz}) = 
\phi(\bfpi^-_{xz},\bfpi^-_{xz}) = \phi(\bfpi^+_{xz},\bfpi^+_{xz}),\\
\phi(\bfe^-_{xy},\bfe^-_{xy}) &=&
\phi(\bfe^+_{xy},\bfe^+_{xy}) ,\\
\phi(\bfpi^-_{xy},\bfpi^-_{xy}) &=& \phi(\bfpi^+_{xy},\bfpi^+_{xy}),\\
\phi(\bfe_x,\bfe_x) &=& \phi(\bfe_y,\bfe_y),\\
\phi(\bfe_z,\bfe_y) &=& \phi(\bfe_z,\bfe_x),\\
\phi(\bfe_y,\bfe_z) &=& \phi(\bfe_x,\bfe_z),\\
\phi(\bfe_y,\bfe_x) &=& \phi(\bfe_x,\bfe_y).
\end{eqnarray*}

When the polarization state of the scattered beam is not measured
we need seven calculations:
\begin{eqnarray*}
	C_0 &=& \frac{2}{9}\big(\sigma(\bfe_x,\bfe_x)+\sigma(\bfe_x,\bfe_y)+\sigma(\bfe_x,\bfe_z)+\sigma(\bfe_z,\bfe_x)\big)
   \\&&+\frac{\sigma(\bfe_z,\bfe_z)}{9}
    ,\\
    C_1 &=& \frac{1}{15}\sigma(\bfe_x,\bfe_x)+\frac{\sigma(\bfe_x,\bfe_y)+\sigma(\bfe_x,\bfe_z)+\sigma(\bfe_z,\bfe_x)}{6}
   \\&&+\frac{\sigma(\bfe_z,\bfe_z)}{30}-\frac{1}{5}\sigma(\bfe^+_{xy},\bfe^+_{xy})
   -\frac{2}{5}\sigma(\bfe^+_{xz},\bfe^+_{xz}).
\end{eqnarray*}
For the full spectra we need nine calculations.
\begin{eqnarray*}
S_0 &=& \frac{2}{3}\sigma(\bfe_x,\bfe_x)+\frac{1}{3}\sigma(\bfe_z,\bfe_z)
-\frac{2}{3}\sigma(\bfpi^+_{xy},\bfpi^-_{xy})
\\&& -\frac{4}{3}\sigma(\bfpi^+_{xz},\bfpi^-_{xz})
+\frac{2}{3}\sigma(\bfe^+_{xy},\bfe^+_{xy})
+\frac{4}{3}\sigma(\bfe^+_{xz},\bfe^+_{xz}),\\
S_1 &=& -2\sigma(\bfe_x,\bfe_x)-2\sigma(\bfe_x,\bfe_y)-2\sigma(\bfe_x,\bfe_z)
\\&& -2 \sigma(\bfe_z,\bfe_x)
-\sigma(\bfe_z,\bfe_z)+\sigma(\bfpi^+_{xy},\bfpi^-_{xy})
\\&& +2\sigma(\bfpi^+_{xz},\bfpi^-_{xz})
+\sigma(\bfe^+_{xy},\bfe^+_{xy})
+2\sigma(\bfe^+_{xz},\bfe^+_{xz}),\\
S_2 &=& -\frac{2}{3}\sigma(\bfe_x,\bfe_x)-\frac{1}{3}\sigma(\bfe_z,\bfe_z)+\frac{5}{3}\sigma(\bfpi^+_{xy},\bfpi^-_{xy})
\\&&
+\frac{10}{3}\sigma(\bfpi^+_{xz},\bfpi^-_{xz})+\frac{1}{3}\sigma(\bfe^+_{xy},\bfe^+_{xy})
+\frac{2}{3}\sigma(\bfe^+_{xz},\bfe^+_{xz}).
\end{eqnarray*}

\subsection{For $T$ or $T_h$}
For the groups $T$ and $T_h$ the relations between 
$\phi(\bfu_l,\bfv_l)$ are
\begin{eqnarray*}
\phi(\bfe^-_{yz},\bfe^-_{yz}) &=& 
\phi(\bfe^+_{yz},\bfe^+_{yz}) = 
\phi(\bfe^-_{xz},\bfe^-_{xz}) = \phi(\bfe^+_{xz},\bfe^+_{xz})\\
&=&\phi(\bfe^-_{xy},\bfe^-_{xy}) =
\phi(\bfe^+_{xy},\bfe^+_{xy}),\\
\phi(\bfpi^-_{yz},\bfpi^-_{yz}) &=& 
\phi(\bfpi^+_{yz},\bfpi^+_{yz}) = 
\phi(\bfpi^-_{xz},\bfpi^-_{xz}) = \phi(\bfpi^+_{xz},\bfpi^+_{xz})\\
&=&\phi(\bfpi^-_{xy},\bfpi^-_{xy}) = \phi(\bfpi^+_{xy},\bfpi^+_{xy}),\\
\phi(\bfe_x,\bfe_x) &=& \phi(\bfe_y,\bfe_y)=\phi(\bfe_z,\bfe_z),\\
\phi(\bfe_x,\bfe_y) &=& \phi(\bfe_y,\bfe_z)=\phi(\bfe_z,\bfe_x),\\
\phi(\bfe_y,\bfe_x) &=& \phi(\bfe_z,\bfe_y)=\phi(\bfe_x,\bfe_z).
\end{eqnarray*}
We need five calculations for the full spectrum
\begin{eqnarray*}
	S_0 &=& \sigma(\bfe_x,\bfe_x) 
	+ 2 \sigma(\bfe^+_{xy},\bfe^+_{xy}) -2 \sigma(\bfpi^+_{xy},\bfpi^-_{xy}),\\
	S_1 &=& -3\sigma(\bfe_x,\bfe_x)-3\sigma(\bfe_x,\bfe_y)-3\sigma(\bfe_y,\bfe_x)
	\\&& +3\sigma(\bfe^+_{xy},\bfe^+_{xy}) 
      +3\sigma(\bfpi^+_{xy},\bfpi^-_{xy}),\\
	S_2 &=& -\sigma(\bfe_x,\bfe_x) 
	+ \sigma(\bfe^+_{xy},\bfe^+_{xy}) +5\sigma(\bfpi^+_{xy},\bfpi^-_{xy}).
\end{eqnarray*}
And we need four calculations if the polarization state of the scattered beam is not measured:
\begin{eqnarray*}
	C_0 &=& \frac{\sigma(\bfe_x,\bfe_x)+\sigma(\bfe_x,\bfe_y)+\sigma(\bfe_y,\bfe_x)}{3},\\
    C_1 &=& \frac{\sigma(\bfe_x,\bfe_x)}{10}+\frac{\sigma(\bfe_x,\bfe_y)+\sigma(\bfe_y,\bfe_x)}{4}-\frac{3\sigma(\bfe^+_{xy},\bfe^+_{xy})}{5}.
\end{eqnarray*}

\begin{figure}[h!]
    \centering
    \includegraphics[width=0.5\linewidth]{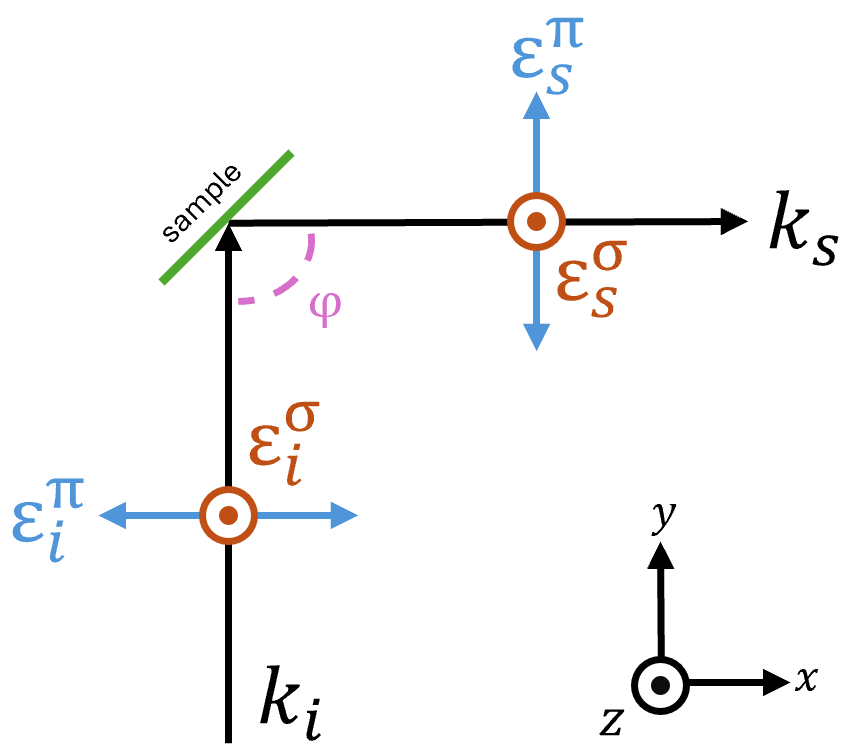}
    \caption{Schematic showing an overview of experiment configuration. Parallel and perpendicular polarization vectors are denoted as $\pi$ and $\sigma$ respectively. The incident and scattered photon momentum are denoted as $\widehat\bfk_i$ and $\widehat\bfk_s$ respectively. The angle of scattered photon detection is given by $\varphi$.}
    \label{fig:ki_ks_scattering}
\end{figure}

\section{Applications}

\subsection{General experiment considerations}


In a RIXS experiment, depending on the energy range of the scattered photon, detection is achieved using an emission spectrometer that can either utilize grazing incidence diffraction gratings for soft X-rays or crystal analyzers for hard X-rays. The sample, crystal analyzer, and detector sit within a Rowland circle.  

With reference to Figure \ref{fig:ki_ks_scattering}, the incoming incident beam is denoted by \ki\ and the emitted fluorescence photons (\ks) that are scattered by the sample are energy discriminated using crystal analyzers, before detection, at an angle $\varphi$ with respect to \ki. The crystal analyzer is usually positioned at 90\textdegree, with respect to the incoming incident beam (\ki), and serves as the reference point from which additional crystal analyzers can be positioned either side. 

The derivations for powder samples from the present paper lead to several general experimental conclusions: 

\begin{itemize}
    \item the RIXS cross-section measured with incident vertical polarization and standard $\varphi$=90\textdegree\ horizontal scattering geometry is identical to the RIXS cross-section measured with incident vertical polarization and back-scattering horizontal geometry.
    \item in hard X-ray RIXS, the use of multiple crystal analyzers leads to the summation of different RIXS spectra unless the each of their intensity is detected individually using a 2D detector. Theoretical modeling should then be performed accordingly using the provided formulas.
    \item for cubic samples, four spectra need to be computed to obtain the correct RIXS cross-section of a powder, contrary to standard XAS where only one spectrum is needed.
\end{itemize}

\begin{figure}[h]
    \centering
    \includegraphics[width=0.75\linewidth]{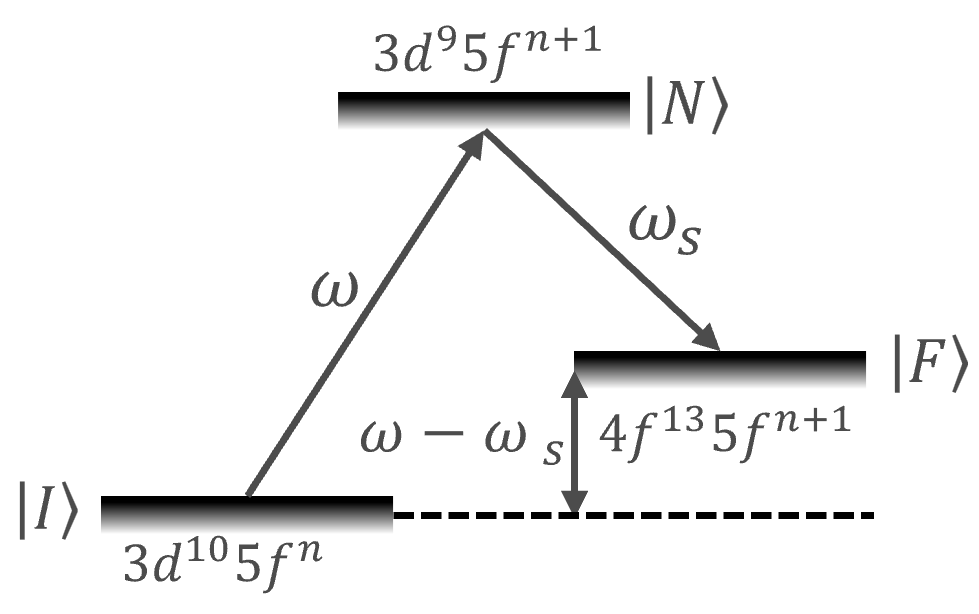}
    \caption{Schematic of \threeDfourF\ RIXS, depicting the correlation between incident ($\omega$), emitted ($\omega_s$), and transferred ($\omega$ - $\omega_s$) energies, and the corresponding electronic configurations of the initial $\lvert  I\rangle$, intermediate $\lvert N\rangle$, and final $\lvert F \rangle$ states.}
    \label{fig:3d4f_RIXS}
\end{figure}

\begin{figure*}[t]
    \centering
    \includegraphics[width=\textwidth]{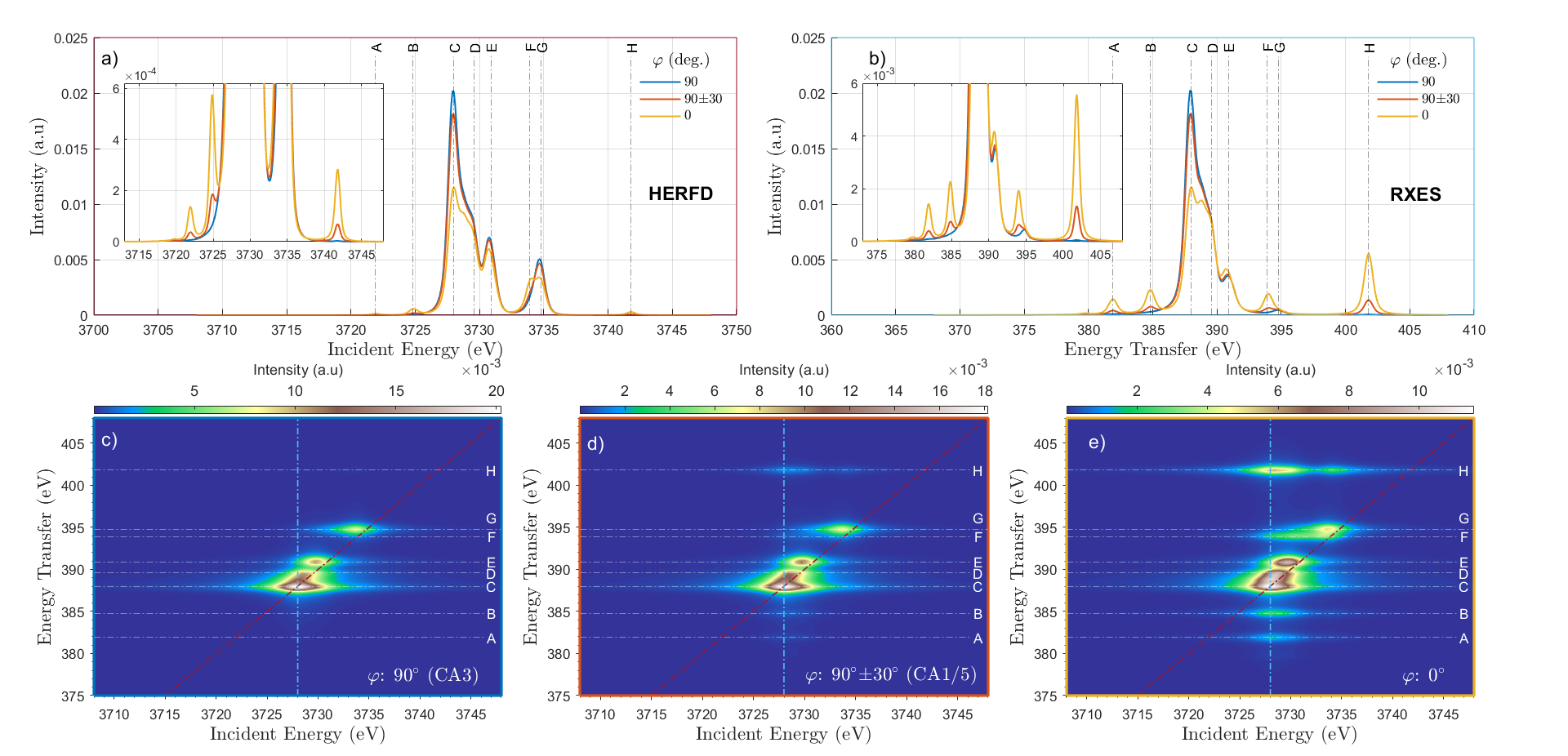}
    \caption{\tetraCl\ a) simulated HERFD cuts through \threeDfourF\ RIXS maps (c-e), inplot is same as (a) but with limited y axis. b) simulated RXES cuts through RIXS maps (c-e), inplot is same as (b) but with limited y axis. c-e) simulated uranium \threeDfourF\ RIXS maps at different crystal analyzer (CA) positions $\varphi$; 90\textdegree, 90\textdegree$\pm$30\textdegree, and back-scattering case of 0\textdegree. HERFD (red dash) and RXES (blue dash) cuts are represented as the diagonal and vertical cuts in the maps. White horizontal lines (c-e) and black vertical lines (a-b) correlate to the most prominent peaks present.}
    \label{fig:RIXS_qCEEcut_qCIEcut}
\end{figure*}

\subsection{Case study: actinide $3d4f$ RIXS}

The formulas derived within section \ref{sympowsect} have been applied to a selection of actinide compounds, previously published and measured using \threeDfourF\ RIXS. It is to be emphasized though, that the derivations are applicable to all RIXS measurements, regardless of which absorption and emission lines are probed.

With respect to actinide X-ray spectroscopy, \threeDfourF\ RIXS is a photon-in photon-out inelastic scattering process of which a \threeD core electron is resonantly excited into the unoccupied valence \fiveF states, while simultaneously detecting fluorescence photons emitted from a \fourF $\rightarrow$ \threeD decay. The complete \threeDfourF\ RIXS process is shown in Figure \ref{fig:3d4f_RIXS}.

Actinide \threeDfourF\ RIXS falls within the tender X-ray region. As such, performing these measurements does not require ultra-high vacuum environments but are yet capable of providing high resolution spectra, exposing spectral features that are otherwise hidden by the large core-hole lifetime broadening if measuring hard X-ray spectroscopy or conventional XAS. This balance of experiment conditions and spectral resolution lends itself as a quintessential technique for probing actinides. Due to this reasoning, \threeDfourF\ RIXS is increasingly being utilized as the technique to probe the valence actinide \fiveF\ orbitals to investigate and gain insights into actinide electronic structure, local coordination, redox, and their subsequent role in covalency \cite{Vitova2010,Kvashnina2013,Kvashnina2017,Amidani2021,Butorin2016,Butorin2016a,Bs2016,Kvashnina2014,Vitova2017,Epifano2019,Burrow2024,Burrow2025,Schacherl2025}.

\begin{figure}[htp]
    \centering
    \includegraphics[width=1\linewidth]{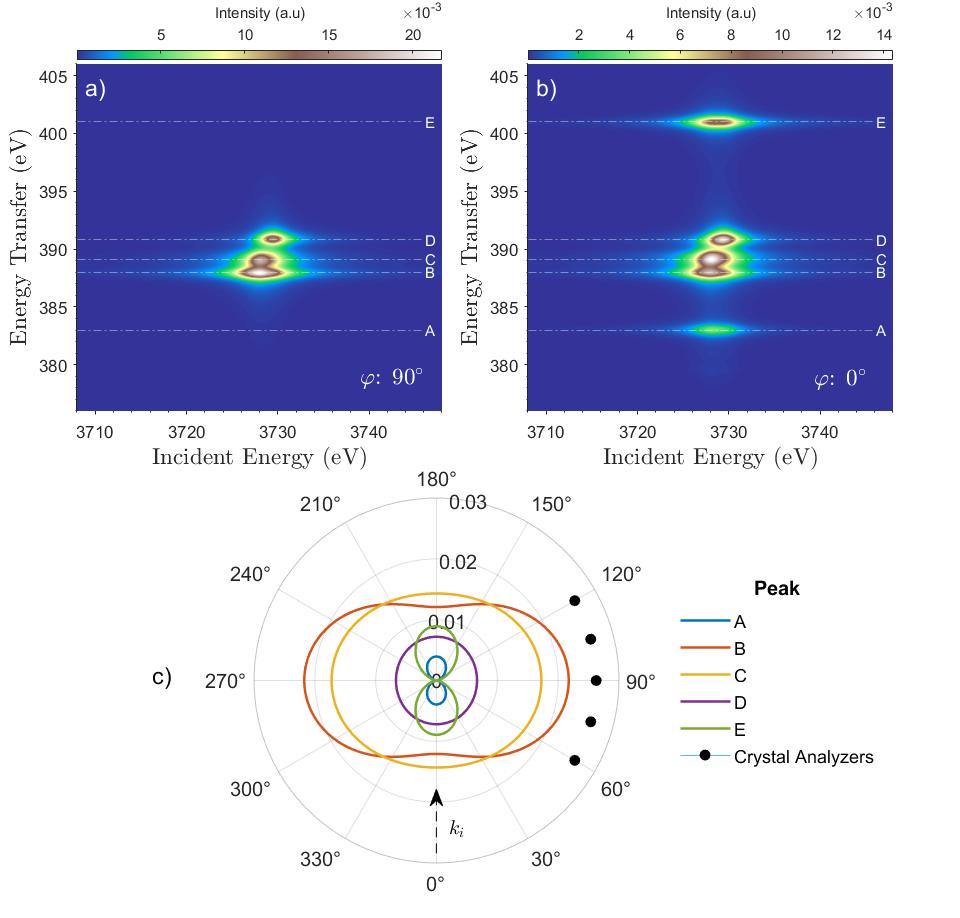}
    \caption{\UOsix\ a-b) simulated uranium \threeDfourF\ RIXS maps at a) $\varphi$=0\textdegree\ and b) $\varphi$=90\textdegree\ with respect to the incident beam, \ki. c) Polar plot of features \textbf{A}-\textbf{E}. The angular axis is representative of $\varphi$ and the radial axis is the peak intensity. Example positions of crystal analyzers are indicated by black dots.} 
    \label{fig:UVIO6_RIXS_0_90_polar}
\end{figure}

\begin{figure}[htp]
    \centering
    \includegraphics[width=1\linewidth]{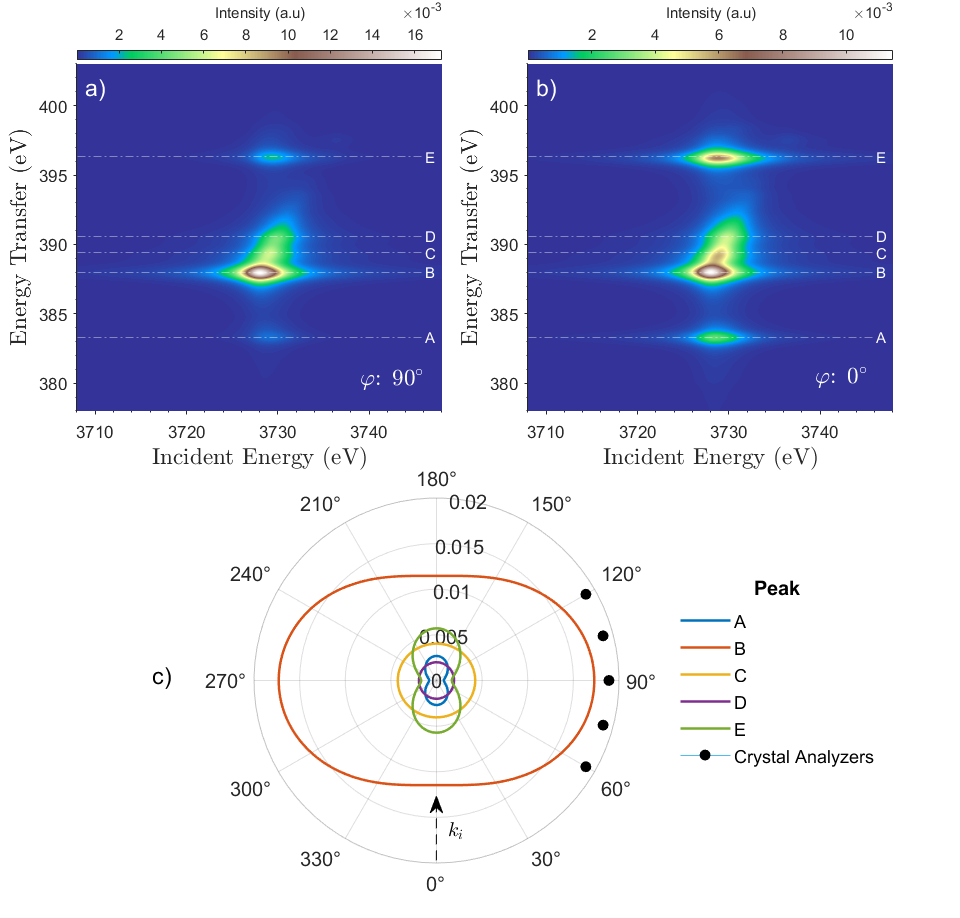}
    \caption{\UOtwo\ a-b) simulated uranium \threeDfourF\ RIXS maps at a) $\varphi$=0\textdegree\ and b) $\varphi$=90\textdegree\ with respect to the incident beam, \ki. c) Polar plot of features \textbf{A}-\textbf{E}. The angular axis is representative of $\varphi$ and the radial axis is the peak intensity. Example positions of crystal analyzers are indicated by black dots.} 
    \label{fig:UIVO2_RIXS_0_90_polar}
\end{figure}

However, due to the relative short divergence of $\varphi$ from the reference crystal analyzers positioned at $\varphi$ = 90\textdegree, we show in Figures \ref{fig:RIXS_qCEEcut_qCIEcut}-\ref{fig:UIVO2_RIXS_0_90_polar} that \threeDfourF\ RIXS measurements performed using these emission spectrometer orientations miss the wealth of information that could otherwise be revealed if the crystal analyzers are positioned towards more extreme back-scattering geometries, ($\varphi$: 90\textdegree $\rightarrow$ 0\textdegree).

To emphasize the contrast of additional information that can be extracted by \threeDfourF\ RIXS, a systematic selection of three uranium compounds, previously published and measured, with common U oxidation states and varying local point-group symmetry have been chosen as case studies, namely; \tetraCl\ ($D_{4h}$)\cite{Amidani2021}, \UOsix\ ($O_h$)\cite{Amidani2021}, and \UOtwo\ ($O_h$)\cite{Butorin2020}, where \ce{U^{VI}} and \ce{U^{IV}} have $5f^{0}$ and $5f^{2}$ ground state electronic structures, respectively. We will consider as typical instrument a five-crystal analyzer spectrometer, as has been developed on several beamlines \cite{Scheinost2021,ablett_multixs_2025,moretti_sala_high-energy-resolution_2018,glatzel_five-analyzer_2021,zimina_cat-actnew_2017}.

In Figure \ref{fig:RIXS_qCEEcut_qCIEcut}(c-d), the simulated RIXS maps correspond to crystal analyzers positioned at $\varphi$=90\textdegree\ and 90\textdegree $\pm$30\textdegree, this is representative of most divergent positions of typical five-analyzers emission spectrometers which typically have crystal analyzers at these positions and  $\varphi$= 90\textdegree $\pm$15\textdegree,  which we do not account for. In order to further highlight the contrast in the possible spectra obtainable, Fig.~\ref{fig:RIXS_qCEEcut_qCIEcut}(e) correlates to {$\varphi$ = 0\textdegree}  and reflects when an emission spectrometer would be orientated in a back-scattering geometry.

From these simulated \threeDfourF\ RIXS maps, constant emission energy (CEE) cuts can be taken (Figure \ref{fig:RIXS_qCEEcut_qCIEcut}(a)), which are representative of experimentally obtained HERFD-XAS spectra, and correspond to the diagonal red-dash cuts through the \threeDfourF\ RIXS maps at the maximum of the N$_6$ ($M_\beta$) emission line. Characteristic spectral features of the CEE cut, peaks \textbf{C-E, G}, are associated to the main whiteline transition and the large \fiveF \DFourH\ ligand field splitting, carrying sensitivity to the incoming incident energy. It is observed that when $\varphi$ is varied from 90\textdegree $\rightarrow$ 0\textdegree, the intensity of these peaks are seemingly not perturbed, with the exception to the whiteline, peak \textbf{C}, which readily decreases in intensity when approaching back-scattering geometries.

Similarly, constant incident energy (CIE) cuts, Figure \ref{fig:RIXS_qCEEcut_qCIEcut}(b), represented as vertical blue-dash cuts through the simulated \threeDfourF\ RIXS maps are representative of resonant X-ray emission spectroscopy (RXES) measurements, carrying sensitivity to the final state of the \threeDfourF\ RIXS process. As the example presented concerns \ce{U^{VI}}, the remaining non-characterized intensity present in the simulated RIXS maps and prominently extracted in the CIE cuts, peaks \textbf{A-B, F, H}, arise from inter-shell $4f^{13}5f^{1}$ spin-exchange interactions, and commonly referred to as satellites. The intensity of these peaks increases drastically as a function of $\varphi$\ (90\ \textdegree$\rightarrow$0\textdegree). It is also noted that the current most divergent crystal analyzer positions ($\varphi$: 90\textdegree$\pm$30\textdegree) of typical five-analyzer spectrometer would be potentially adequate to capture partial intensity of these features in experiment conditions, depending on signal-noise and other experiment considerations.

Upon comparing simulations presented here using the derivations of section \ref{sympowsect} with the simulations previously published ~\cite{Amidani2021}, some key points are made. The authors previously modeled their experimental HERFD-XAS spectra of \tetraCl\ using an X-ray absorption ligand field multiplet model that considers the \fiveF\ ligand field splitting, $3d^{9}5f^{1}$ repulsion and spin-exchange interactions, and spin-orbit coupling. Their best simulation of the \tetraCl\ HERFD data most noticeably required a scaling the \threeDdashfiveF\ Slater integrals to 50\% of their atomic values. Conversely, what we present is that by employing a ligand field multiplet model that encompasses the entire \threeDfourF\ RIXS process, implying additional consideration in the simulations to include the $4f^{13}5f^{1}$ repulsion and spin-exchange interactions and their respective spin-orbit coupling constants, and that of which employs the equations for \DFourH\ in section \ref{sympowsect}, we are able to come close to reproducing the fine structure in their spectra. The key factor though is that in the simulations we present, the Slater integrals were solely scaled, conventionally, to 80\% of their atomic value~\cite{Lynch1987}, with no other scaling required. Solely by accounting for the position of the crystal analyzers with respect to $\varphi$ and the inclusion of our derivations when calculating the \threeDfourF\ RIXS spectra, we are, in general, able to now provide a method that more accurately simulates experimental data.


Maintaining a uranium $5f^{0}$ electronic configuration but changing the symmetry from \tetraCl\ \DFourH\ to \UOsix\ \Oh, the results are similar in nature. Figure \ref{fig:UVIO6_RIXS_0_90_polar}(a-b) show simulated \threeDfourF\ RIXS maps of \UOsix\ in both extreme, perpendicular ($\varphi$=90\textdegree) and back-scattering ($\varphi$=0\textdegree), geometries. It is clear that the experiment orientation has a limited sensitivity to peaks \textbf{B}-\textbf{D}, and can be attributed to the \fiveF ligand field splitting. Contrarily, peaks \textbf{A} \& \textbf{E} carry significant sensitivity to the \fourFdashfiveF\ inter-shell spin-exchange interactions in the final state of \threeDfourF\ RIXS. Another perspective of analyzing the sensitivity of peak intensity, peaks \textbf{A}-\textbf{E}, as a function of $\varphi$ is shown in Figure \ref{fig:UVIO6_RIXS_0_90_polar}(c). The intensity of the peaks is shown on the radial axis of the polar plot, while the angular axis shows $\varphi$, typical five-crystal analyzer $\varphi$ positions  are shown as a reference emission spectrometer configuration. When measuring perpendicular to the incident beam ($\varphi$ = 90\textdegree), the intensity of peaks \textbf{A} and \textbf{E} approach 0 and would not be observed experimentally, whilst all other features are at their maxima. Conversely, at $\varphi$ = 0\textdegree, this relation is inverted such that the satellite intensities are at their respective maxima and peaks \textbf{B}-\textbf{D} are at their lowest. 

Maintaining \Oh\ symmetry but diverging from \ce{U^{VI}} $5f^0$ (\UOsix) to \ce{U^{IV}} $5f^2$ (\UOtwo) ground state electronic configuration subsequently results in Figure~\ref{fig:UIVO2_RIXS_0_90_polar}. Prior to characterizing the features present in the simulations, starting from a $5f^{2}$ ground state electronic structure, this subsequently leads to an intermediate $\lvert$N$\rangle$ $3d^{9}5f^{3}$ and final $\lvert$F$\rangle$ $4f^{13}5f^{3}$ state electronic configuration during the RIXS process. With particular emphasis to the final state electronic configuration, it consequently leads to necessity to consider \fourFdashfiveF\ and \fiveFdashfiveF\ inter-shell spin-exchange and repulsion interactions. Comparing between Figures \ref{fig:UVIO6_RIXS_0_90_polar}(c) and \ref{fig:UIVO2_RIXS_0_90_polar}(c), where the experiment geometry is perpendicular to the incoming beam ($\varphi$ = 90\textdegree), peaks \textbf{A} and \textbf{E} would each now have observable intensity in \ce{U^{IV}} experimental \threeDfourF\ RIXS measurements. The same observations as before, with reference to Figure~\ref{fig:UIVO2_RIXS_0_90_polar}(c), are observed when the crystal analyzers are orientated perpendicular to the incoming beam ($\varphi$ = 90\textdegree), the main whiteline intensity and features \textbf{B}-\textbf{D} arising from the \fiveF ligand field splitting are most intense. Additionally secluded within these peaks, however, is intensity that originates from  \fiveFdashfiveF\ repulsion interactions, although it is proposed that $\varphi$ carries minor influence on the intensity of these features. Although, when approaching back-scattering geometry ($\varphi$: 90\textdegree$\rightarrow$0\textdegree), the satellites corresponding to inter-shell \fourFdashfiveF\ spin-exchange interaction are most prominent, following the same trends as \ce{U^{VI}}.

As previous authors have used the energy position and intensities of the satellites to gain information into the effective value of Slater integrals to further derive insights into chemical bonding, we show however that the implication from these examples, such satellites present in \threeDfourF\ RIXS, and others, can serve as an anchor for determining the correct scaling of Slater integrals, as solely the intensity of these features are perturbed as function of $\varphi$, not their energetic positioning, which only changes as a consequence of scaling.

To further elaborate on the sensitivity of \threeDfourF\ RIXS to inter-shell \fiveFdashfiveF\ repulsion and \fourFdashfiveF\ spin-exchange interactions, a series of \ce{An^{IV}O2} (An = Th, Np, Pu) in cubic symmetry (\Oh) as previously published\cite{Butorin2020} have been simulated as example cases where the ground state \fiveF electron occupancy changes from $5f^{0}$ to $5f^{3}$ to $5f^{4}$ respectively. These can be found in the supplementary information.

\section{Conclusion and prospects}

In this work, we provided a general method —applicable to all point group symmetries— to express a RIXS spectrum in terms of its fundamental spectra. In other words, we determined the minimal number of spectra that must be measured or calculated in order to fully access all the spectral information contained in a sample. The analysis was carried out within the electric dipole approximation for both the incoming and outgoing photons.

A direct application of this approach is the derivation of simple and practical expressions for the RIXS spectrum of isotropic samples with the most common sample symmetries (cubic, tetrahedral, hexagonal, trigonal, and tetragonal). We expect that these easy-to-use formulas will stimulate new experimental measurements, in particular by exploring the dependence on the scattering angle and the polarization of the outgoing beam, which for example could help to enhance elusive spectral features by choosing more appropriate experimental configurations. Accounting for this dependence may also lead to a re-examination, and possibly a reinterpretation, of existing data in the literature.

To illustrate the derivations, we show that there is a significant angular dependence in actinide \threeDfourF\ RIXS by simulating previously reported literature. In each of these examples, it is shown that final-state RIXS satellite intensities are significantly perturbed as a function of the spectrometer geometry. As actinide X-ray measurements are performed on isotropic powder samples, we have shown that it is crucial to use the formulas presented within the present paper to correctly simulate satellite intensities and subsequently obtain the correct effective scaling for Slater integrals. Furthermore, from an experiment perspective, we highlight that ideal experiment conditions should use a 2D detector for RIXS measurements. This would facilitate spectrometer designs to have a large divergence with respect to \ks, and allow for the individual analysis of the scattered photon intensity from each crystal analyzer. 

The present work can also be used to determine the minimum number of experimental spectra that must be measured to extract all the information available from RIXS spectroscopy.

A natural extension of this work, which is currently in progress, concerns the case of an electric quadrupole approximation for the incoming photon combined with an electric dipole approximation for the outgoing photon. This situation is relevant to many experiments on transition metals. Another perspective is to include the dipole–quadrupole interference term for the incoming photon, which gives rise to the XNCD effect in RIXS.

\section{Acknowledgements}
We are very grateful to Coraline Letouz{\'e} for fruitful discussions and comments.
After completion of this work, Tagliavini and coll.~\cite{Tagliavini-25}
published an article on similar subjects. 

\appendix
\section{Ligand field multiplet simulations}\label{methods:LFM}

Multiplet simulations of the \threeDfourF\ RIXS planes were carried out using Quanty Version 0.7 beta \cite{QHaverkort2016}. To emphasize the features, a slightly narrower broadening was used than that previously published, such that a Lorentzian broadening of 3 eV and 0.5 eV for the $3d^{9}$ and $4f^{13}$ core-hole lifetimes, and a Gaussian broadening of 0.5 eV, associated to experimental resolution. As Quanty is semi-empirical, a modular Hamiltonian was be constructed consisting of the \fiveF ligand field splitting, electron-electron interactions, and spin-orbit coupling in the initial, intermediate, and final electronic configurations of \threeDfourF\ RIXS. A small magnetic field was applied on the Z axis. The on-site energy of regarding the \fiveF ligand field splitting were the same as previously reported \cite{Amidani2021,Butorin2020}. All electron-electron interactions, associated to the Coulomb inter-shell repulsion \& spin-exchange interactions, were scaled at 80\% of the Hartree-Fock value. Spin-orbit coupling constants were kept at a scaling of 100\%. No scaling has been applied regarding the intensity of the simulated RIXS maps. \\ 

\bibliographystyle{apsrev4-2-titles}
\bibliography{qed,biblio_MOJY}
\end{document}


\title{Supplementary Information: Angular dependence and powder average of resonant inelastic x-ray scattering}

\author{Myrtille O. J. Y Hunault}
\author{Timothy G. Burrow} 
\affiliation{Synchrotron SOLEIL, L'Orme des Merisiers, 91190 Saint-Aubin, France}

\author{Fabien Besnard}
\affiliation{11 all\'ee Hector Berlioz, 95230 Soisy-sous-Montmorency, France}

\author{Am\'elie Juhin}
\email[Corresponding author: ] {amelie.juhin@upmc.fr}
\author{Christian Brouder}
\affiliation{Sorbonne Universit\'e, Mus\'eum National d'Histoire Naturelle, 
	UMR CNRS 7590, Institut de Min\'eralogie, de Physique des Mat\'eriaux et 
	de Cosmochimie, IMPMC, 75005 Paris, France}

\maketitle

\section{Ligand field multiplet simulations}

Presented are $3d4f$ RIXS simulations, using the RIXS cross-section formulae provided within the main text, of \ce{An^{IV}O2} (An  = Th, Np, Pu) as previously published \cite{Butorin2020}. The ground state $5f$ electron occupancy is $5f^{n}$, $n$ = 0, 3, and 4, respectively. Ligand field multiplet calculations were completed using the methodology outlined in the appendix of the main manuscript.

\begin{figure}[htp]
    \centering
    \includegraphics[width=0.6\linewidth]{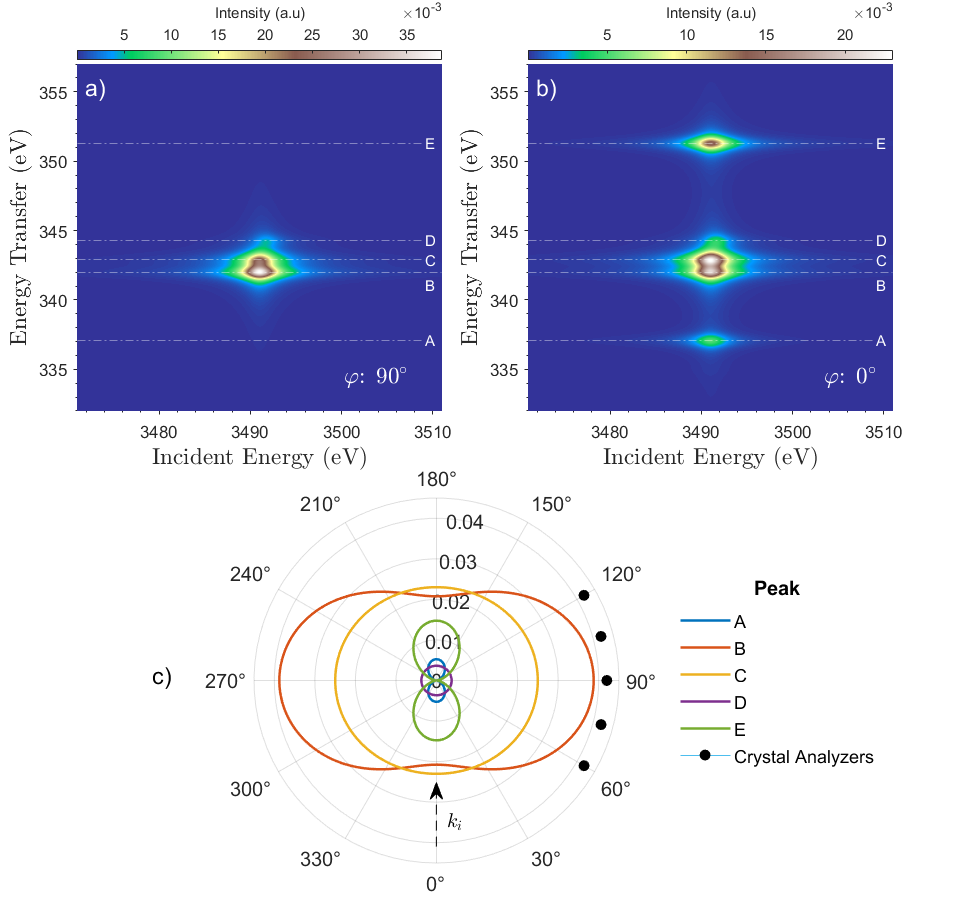}
    \caption{\ce{Th^{IV}O2}\ a-b) simulated thorium $3d4f$ RIXS maps at a) $\varphi$=0\textdegree\ and b) $\varphi$=90\textdegree\ with respect to the incident beam, \ki. c) Polar plot of features \textbf{A}-\textbf{E}. The angular axis is representative of $\varphi$ and the radial axis is the peak intensity. Example positions of crystal analyzers are indicated by black dots.} 
    \label{figSI:ThIVO2_RIXS_0_90_polar}
\end{figure}

\begin{figure}[htp]
    \centering
    \includegraphics[width=0.6\linewidth]{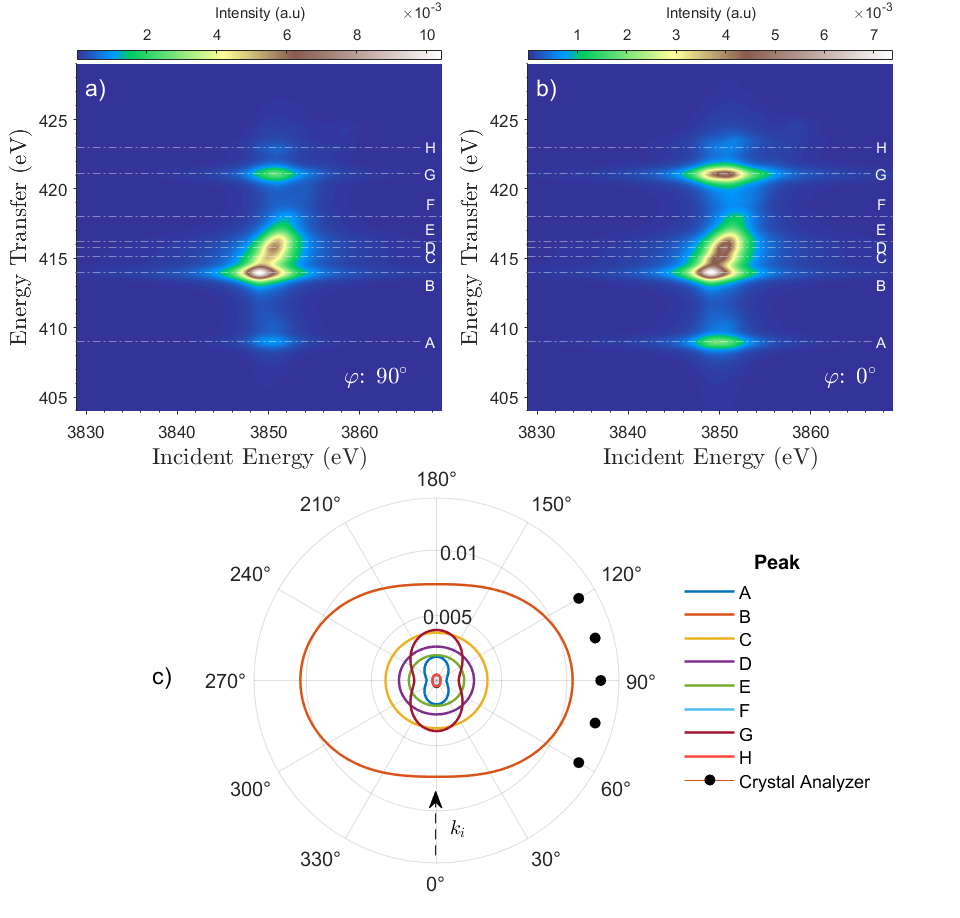}
    \caption{\ce{Np^{IV}O2}\ a-b) simulated neptunium $3d4f$ RIXS maps at a) $\varphi$=0\textdegree\ and b) $\varphi$=90\textdegree\ with respect to the incident beam, \ki. c) Polar plot of features \textbf{A}-\textbf{E}. The angular axis is representative of $\varphi$ and the radial axis is the peak intensity. Example positions of crystal analyzers are indicated by black dots.} 
    \label{figSI:NpIVO2_RIXS_0_90_polar}
\end{figure}

\begin{figure}[htp]
    \centering
    \includegraphics[width=0.6\linewidth]{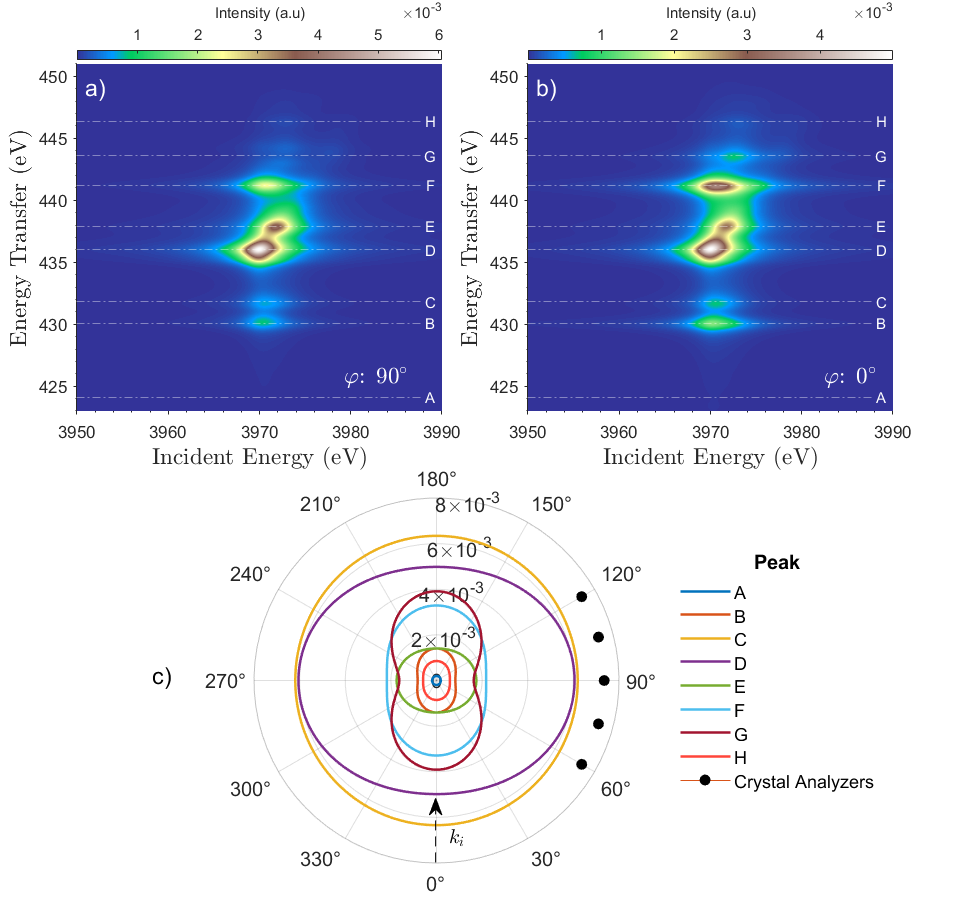}
    \caption{\ce{Pu^{IV}O2}\ a-b) simulated plutonium $3d4f$ RIXS maps at a) $\varphi$=0\textdegree\ and b) $\varphi$=90\textdegree\ with respect to the incident beam, $\widehat\bfk_i$. c) Polar plot of features \textbf{A}-\textbf{E}. The angular axis is representative of $\varphi$ and the radial axis is the peak intensity. Note for clarity, intensity of peaks (\textbf{A},\textbf{C},\textbf{G},\textbf{H}) were increased by x10 (c). Example positions of crystal analyzers are indicated by black dots.} 
    \label{figSI:PuIVO2_RIXS_0_90_polar}
\end{figure}

\bibliographystyle{apsrev4-2-titles}
\bibliography{qed,biblio_MOJY}